\documentstyle[12pt,epsf]{article}
\newcommand{\lsim}{\mathrel{\raisebox{-.6ex}{$\stackrel{\textstyle<}{\sim}$}}}
\newcommand{\gsim}{\mathrel{\raisebox{-.6ex}{$\stackrel{\textstyle>}{\sim}$}}}
\def\eps{\epsilon}
\def\ts{\textstyle}
\def\nue{$\nu_e$}
\def\numu{$\nu_\mu$}
\def\nutau{$\nu_\tau$}
\def\nus{$\nu_s$}
\def\nuzero{$\nu_0$}
\def\nuone{$\nu_1$}
\def\nutwo{$\nu_2$}
\def\nuthree{$\nu_3$}

\def\beq{\begin{equation}}
\def\eeq{\end{equation}}
\def\bea{\begin{eqnarray}}
\def\eea{\end{eqnarray}}

\begin{document}

\thispagestyle{empty}

\font\fortssbx=cmssbx10 scaled \magstep2
\hbox to \hsize{
\hbox{\fortssbx University of Wisconsin - Madison}
      \hfill$\vcenter{
\hbox{\bf MADPH-98-1054}
\hbox{\bf HAWAII-511-900-98}
\hbox{\bf VAND-TH-98-05}
\hbox{\bf AMES-HET-98-08}
       \hbox{June 1998}}$ }

\vspace{.5in}

\begin{center}
{\bf VARIATIONS ON FOUR--NEUTRINO OSCILLATIONS}
\\
\vskip 0.7cm
{V. Barger$^1$, S. Pakvasa$^2$, T.J. Weiler$^3$, and K. Whisnant$^4$}
\\[.2cm]
$^1${\it Department of Physics, University of Wisconsin, Madison, WI
53706, USA}\\
$^2${\it Department of Physics and Astronomy, University of Hawaii,
Manoa, HI 96822, USA}\\
$^3${\it Department of Physics and Astronomy, Vanderbilt University,
Nashville, TN 37235, USA}\\
$^4${\it Department of Physics and Astronomy, Iowa State University,
Ames, IA 50011, USA}\\
\end{center}

\smallskip

\begin{abstract}

We make a model--independent analysis of all available data that indicate
neutrino oscillations. Using probability diagrams,
we confirm that a mass spectrum with two nearly degenerate pairs of
neutrinos separated by a mass gap of $\simeq1$~eV is preferred over
a spectrum with one mass eigenstate separated from the others.
We derive some new relations among
the four--neutrino mixing matrix elements.
We design four-neutrino mass matrices with three active neutrinos
and one sterile neutrino that naturally incorporate maximal
oscillations of atmospheric $\nu_\mu$ and explain the solar neutrino
and LSND results. The models allow either a large or small angle
MSW or vacuum oscillation
description of the solar neutrino deficit. The models predict
(i) oscillations of either $\nu_e \rightarrow \nu_\tau$
or $\nu_e \rightarrow\nu_s$ in long--baseline experiments at $L/E \gg
1$~km/GeV, with amplitude determined by the LSND oscillation amplitude
and argument given by the atmospheric $\delta m^2$, and
(ii) the equality of the $\nu_e$
disappearance probability, the $\nu_\mu$ disappearance probability, and
the LSND $\nu_\mu\rightarrow\nu_e$ appearance probability in
short--baseline experiments.


\end{abstract}

\thispagestyle{empty}
\newpage

\section{Introduction}

The long--standing solar neutrino deficit~\cite{SSM,solar}, the
atmospheric neutrino anomaly~\cite{atmos,SuperK,SKam:up/down,oldatmos},
and the results from the LSND experiment on ${\bar \nu}_e$ neutrinos
from $\mu^+$ decay and \nue\ neutrinos from $\pi^+$ decay~\cite{LSND}
can each be understood in terms of oscillations between two neutrino
species~\cite{review}.  Interestingly, the solar, atmospheric, and
terrestrial (LSND) neutrino oscillations have different $L/E$ and
therefore require different neutrino mass--squared differences $\delta
m^2$ to properly describe all features of the data.  For example, if the
atmospheric and LSND $\delta m^2$ scales are the same~\cite{sunalone},
one forfeits the recently reported zenith-angle dependence and up/down
asymmetry of the atmospheric neutrino flux \cite{SuperK,SKam:up/down}.
Alternatively, if the solar and atmospheric $\delta m^2$
scales~\cite{LSNDalone} are the same, the reduction in the solar
neutrino flux is energy-independent, contrary to the three solar
experiments which infer different oscillation probabilities in different
neutrino energy regions~\cite{krastev1}. Since three distinct
mass-squared differences cannot be constructed from just three neutrino
masses, the collective data thus argue provocatively for more than
three oscillating flavors. An alternative but less compelling
possibility is to introduce new lepton--flavor changing operators
with coefficients small enough to evade present exclusion limits, but
large enough to explain the small LSND amplitude~\cite{jm98}.


If all of the existing observations are confirmed, a viable solution is
to invoke one or more additional species of sterile light
neutrino~\cite{VB80}, thereby introducing another independent mass scale
to the theory.  The additional neutrino must be sterile, i.e. without
Standard Model gauge interactions, to be consistent with LEP
measurements of $Z \rightarrow \nu\bar\nu$~\cite{Znunubar}.  The
introduction of a sterile neutrino to complement the three active
neutrinos has had some phenomenological success~\cite{models}.

In this paper we propose and study mass matrices for four--neutrino
models (three active plus one sterile)
that can accommodate all the present data.  Once a fourth neutrino is
admitted to the spectrum, it is no longer mandatory that the $\nu_\mu$
mix with the $\nu_\tau$ at the atmospheric scale.  The $\nu_\mu$ may
instead mix with the sterile $\nu_s$, or with some linear combination of
$\nu_s$ and $\nu_\tau$.  Similarly, the $\nu_e$ may mix with a linear
combination of $\nu_s$ and $\nu_\tau$.

At first sight the mixing of a sterile neutrino with active flavor neutrinos
seems to be stringently constrained by Big Bang nucleosynthesis
(BBN) physics.  The bound
\begin{equation}
\delta m^2 \sin^22\theta < 10^{-7} {\rm~eV}^2
\label{bbn}
\end{equation}
on the mass-squared difference $\delta m^2$ and the mixing angle of the
sterile neutrino was inferred to avoid thermal overpopulation of the
``extra'' sterile neutrino species\cite{BBN}.  However, there are
significant caveats to this bound. One is the fact that some recent
estimates of $N_\nu$ using higher abundances of $^4$He yield
considerably weaker bounds \cite{He4}. Another is that a small asymmetry
$(n_\nu - n_{\bar\nu})/n_\gamma \gsim 7\times10^{-5}$ of flavor
neutrinos (but large compared to the present baryon asymmetry $\Delta
n_B/n_\gamma \sim 10^{-10}$) at $t > 0.1$~s is enough to suppress
$\nu_\mu-\nu_s$ oscillations and then the bound of Eq.~(\ref{bbn}) does
not apply~\cite{suppress}. Such asymmetries, in fact, can be generated
with the kind of model parameters considered herein (as shown in
Ref.~\cite{fv97,ls97}).  In light of this observation that BBN may allow
sizeable mixing between sterile and active neutrinos, we consider both
the small and large mixing with sterile neutrinos in this work.

We review all existing data that indicate neutrino oscillations, and then
perform a model--independent analysis of the data using four--neutrino
unitarity constraints.  A very useful tool for this unitarity analysis
is the set of probability rectangles, which we explain and exploit. We
draw several model--independent conclusions for the four--neutrino
universe.

We design a five--parameter neutrino mass matrix which can account
for each of the three viable solar solutions and accommodate the
atmospheric and LSND observations.  The three solar possibilities are the
small-angle matter-enhanced (SAM)~\cite{MSW1,MSW2,MSW3,MSW4},
large-angle matter
enhanced (LAM)~\cite{lam} and large--angle vacuum long--wavelength
(VLW)~\cite{bpw81,justso} explanations of the solar
neutrino deficit.  Our mass matrix yields maximal oscillations of
atmospheric $\nu_\mu$. We consider the possibility
that the solar data is explained by $\nu_e\rightarrow\nu_s$ or
$\nu_e\rightarrow\nu_\tau$ oscillations, in which case the atmospheric
neutrino data is explained by either $\nu_\mu\rightarrow\nu_\tau$ or
$\nu_\mu\rightarrow\nu_s$ oscillations, respectively. We also
consider the possibility that both atmospheric and solar neutrino
oscillations have $\nu_s$ and $\nu_\tau$ components. Lack of
$\nu_s$--$\nu_\tau$ discrimination in the present data is the major
source of ambiguity in the four--neutrino model. We discuss how
future experiments can resolve this ambiguity.

In Sec.~2 we summarize the oscillation probability formulas and utilize a
probability formalism, based on unitarity of the mixing matrix, which
permits a simple visual represention of mixing. In Sec.~3 we begin with
a brief discussion of the three classes of experiments and the neutrino
mass and mixing parameters needed to explain them.
We then use probability rectangles to display the inferences
from the data for any four--neutrino scheme.  In Sec.~4 we employ the
probability rectangles to argue against a neutrino mass spectrum with
one eigenstate separated from three other nearly--degenerate states
(which we will refer to as the 1+3 spectrum) in favor of two nearly
degenerate mass pairs (which we will refer to as the 2+2 spectrum). We
also derive some new relations among elements of the mixing matrix that
result from data and unitarity which are satisfied in a four--neutrino
model for certain ranges of the parameters.
Then in Sec.~5 we present a mass matrix whose
eigenvalues consist of a nearly degenerate neutrino pair at $\sim
1.4$~eV and another nearly degenerate pair at low mass, as illustrated
in Fig.~1. We show how the existing data almost uniquely fixes the model
parameters (once a solar scenario is specified) and strictly determines
what new phenomenology the model predicts.  In Sec.~6 we derive
expressions for the oscillation probabilities in our models in terms of
the current neutrino experimental observables. We present the model
predictions in Sec.~7.  The new observable signature for the model
is $\nu_e \leftrightarrow \nu_\tau$ or $\nu_e \leftrightarrow \nu_s$
oscillations for $L/E \gg 1$~km/GeV, depending on whether the atmospheric
oscillations are $\nu_\mu\rightarrow\nu_\tau$ or
$\nu_\mu\rightarrow\nu_s$, respectively.  Section~8 contains some
discussion, and a summary.

\section{Formalism}

\subsection{Oscillation amplitudes}

To simplify the analysis of the available data,
we will ignore possible CP violation
and work with a real--valued mixing matrix $U$.
Accordingly, the general formula for the vacuum oscillation probabilities
becomes~\cite{VBreal}
\begin{equation}
P(\nu_\alpha\to \nu_\beta) = \delta_{\alpha\beta}
- 4 \sum_{k<j} U_{\alpha k} U_{\beta k}
U_{\alpha j} U_{\beta j} \sin^2 \Delta_{jk} \,,
\label{oscprob}
\end{equation}
where $\Delta_{jk} \equiv \delta m_{jk}^2 \,L/4E = 1.27
(\delta m^2_{jk}/{\rm eV}^2) (L/{\rm km})/(E/{\rm GeV})$,
$\delta m^2_{jk}\equiv m^2_j -m^2_k$, and the sum is over all
$j$ and $k$ subject to $k<j$.

For oscillations of two neutrinos, the oscillation amplitude (i.e., the
coefficient of the $\sin^2\Delta_{jk}$ term) is given by $\sin^22\theta$,
where $\theta$ is the mixing angle between the two neutrino states.
More generally for an arbitrary number of neutrinos, the amplitude of
the $\nu_\alpha$ to $\nu_\beta$ oscillation in the absence of CP
violation is seen to be
\beq
A^{\alpha\beta}=-4\sum_{k<j}
U_{\alpha j}U_{\beta j}U_{\alpha k}U_{\beta k}\,,\qquad \alpha\ne\beta\,,
\label{amp}
\eeq
where the sum is over mass states with mass-squared differences
appropriate for the $L/E$ of the particular experiment.  We note that
the oscillation amplitudes defined here are only for those oscillations
at a particular $\Delta$ scale in Eq.~(\ref{oscprob}). We will use
subscript labels on the amplitude to identify the $\Delta$ scale (which
is determined by the relevant $\delta m^2$ and $L/E$) for the particular
experiment: ``sbl'' will denote short--baseline experiments such as
LSND, ``atm'' will denote atmospheric and long--baseline experiments,
and ``sun'' will denote extraterrestrial experiments, especially those
with solar neutrinos.

We will use superscripts on the amplitude to identify the oscillation
flavors, unless it is obvious from the context; in the absence of CP
violation, $A^{\alpha\beta}=A^{\beta\alpha}$. With four neutrino
states, $U$ is a $4\times 4$ mixing matrix.
We also define the amplitude for $\nu_\alpha$ disappearance
\beq
A^{\alpha\not\alpha} \equiv \sum_{\beta\ne\alpha} A^{\alpha\beta} \,,
\label{ampnot}
\eeq
where ${\not\alpha}$ represents a sum over neutrino
flavor eigenstates other than $\nu_\alpha$.  The mixing--matrix
elements $U$, and therefore the amplitudes $A$, depend on the
environment, e.g., matter vs. vacuum.  Throughout this paper we will
quote values for the oscillation amplitudes {\it in vacuum}.

\subsection{Probability rectangles and a theorem}

The ``probability rectangles'' used by
Liu and Smirnov~\cite{ls97} visually illustrate the mixing of the
flavor eigenstates among the mass eigenstates.  To construct the
probability rectangles, we introduce the notation
\beq
P_{\alpha j} \equiv |U_{\alpha j}|^2,
\label{prob}
\eeq
such that $P_{\alpha j}$ is the probability that the $\alpha^{th}$ flavor
state is found in the $j^{th}$ mass state, or, alternatively, the
probability that the $j^{th}$ mass state is contained in the
$\alpha^{th}$ flavor state.  Therefore, when CP--violation is neglected,
the real mixing--matrix elements are determined by the probabilities up
to a sign: $U_{\alpha j}=\pm\sqrt{P_{\alpha j}}$.  In principle, these
signs may be determined by arranging for orthogonality of the rows, and
columns, in the unitary mixing matrix $U$.

By unitarity of $U$ we have
\beq
\sum_{\alpha} P_{\alpha j} =1
\label{uni1}
\eeq
for each mass state $j$, and
\beq
\sum_j P_{\alpha j} =1
\label{uni2}
\eeq
for each flavor state $\alpha$.  Thus, if each mass state is represented
as a rectangle of unit area, then the fractional area assigned to
$P_{\alpha j}$ within the rectangle is a graphical representation of the
value of $P_{\alpha j}$.  The probabilities $P_{\alpha j}$ depend on
whether the environment is vacuum or matter.  For consistency, we will
always display vacuum probabilities in the rectangles.  When the
probability rectangles are displayed along a vertical axis labeled with
mass--squared, the $\delta m^2$ values relevant for the various
experiments are readily visualized. Figure~2 gives an example of the
probability rectangles for a four--neutrino model. An inverted 2+2
mass spectrum, where the solar $\nu_e$ oscillation is
driven by the separation of the heavier two states and the atmospheric
$\nu_\mu$ oscillation is driven by the separation of the lighter two
states, may also be possible, but is not considered.

The following mini-theorem will prove to be useful:\\ {\it In the
absence of matter effects, the amplitude $A^{\alpha\not\alpha}$ is
independent of how the $P_{\not\alpha j}$ probabilities are partitioned
among the mass eigenstates.}\\ The proof of this statement relies on the
insertion of $\sum_{\beta\neq\alpha} U_{\beta j} U_{\beta k} = -
U_{\alpha j} U_{\alpha k}$ into\\ $A^{\alpha\not\alpha}=
-4\sum_{\beta\neq\alpha}\sum_{j, k> j} U_{\alpha j} U_{\alpha k}
U_{\beta j} U_{\beta k}$, to get
\beq
A^{\alpha\not\alpha}=4\sum_{j>k} P_{\alpha j} P_{\alpha k},
\label{theorem}
\eeq
where, as in Eq.~(\ref{amp}), the sum in Eq.~(\ref{theorem}) is over
all mass states with mass-squared differences appropriate for the
$L/E$ of the particular experiment. In Eq.~(\ref{theorem}),
$A_{\alpha\not\alpha}$ is manifestly independent of the partitioning
of the $P_{\not\alpha j}$ probabilities since it involves only
the $P_{\alpha j}$. This theorem demonstrates the
limitations on information derivable from disappearance experiments.

The minitheorem fails in the presence of matter effects because the
partitioning of the flavor probabilities including $P_{\alpha j}$ are
altered. That is, with matter effects the amplitude
$A_{\alpha\rightarrow\not\alpha}$ {\it does} depend on how the
$P_{\not\alpha j}$ are partitioned among the mass states. Matter will
also alter the oscillation wavelength, causing further changes in the
phenomenology of experiments sensitive to the oscillations rather than
their averages. Matter effects have the potential to resolve the
$\nu_s$--$\nu_{\tau}$ ambiguity, as do some other measurements.  We
discuss these possibilities in Sec.~7.

\section{Experimental constraints}

\subsection{Short baseline: LSND, reactors, and accelerators}

The LSND experiment~\cite{LSND} reports positive appearance results for
$\bar\nu_\mu \rightarrow \bar\nu_e$ oscillations from $\mu^+$ decay at
rest (DAR) and for $\nu_\mu\rightarrow\nu_e$ oscillations from $\pi^+$
decay in flight (DIF). The DAR data has higher statistics, but the
allowed regions for the two processes are in good agreement. There are
also restrictions from the null results of the BNL E-776~\cite{E-776}
and KARMEN~\cite{KARMEN} $\nu_\mu\rightarrow\nu_e$ oscillation search
experiments. The combined data suggest $\nu_\mu\rightarrow\nu_e$ vacuum
oscillation parameters that lie approximately along the line segment
described by
\begin{equation}
0.3 {\rm~eV}^2 \le \delta m^2_{sbl} =
{0.030 {\rm~eV}^2 \over (A^{\mu e}_{sbl})^{0.7}} \le 2.0 {\rm~eV}^2 \;.
\label{lsnddata}
\end{equation}
However, values for $\delta m^2_{sbl}$ as high as 10~eV$^2$ are also
allowed for $A^{\mu e}_{sbl} \simeq .0025$, although
values above 3~eV$^2$ are disfavored by the r--process mechanism
of heavy element nucleosynthesis in supernovae~\cite{rprocess}.

There are also relevant data from the Bugey reactor experiment which
searches for $\bar\nu_e$ disappearance~\cite{Bugey},
and from the CDHS\cite{CDHS} and CCFR\cite{CCFR} experiments which set
bounds on \numu\ disappearance.

The combined short baseline data set
for $A^{e\not e}$, $A^{\mu\not\mu}$, and $A^{\mu e}$ will be
used in Sec.~4 to argue against a  hierarchical neutrino mass spectrum
in favor of two pairs of nearly degenerate masses in the four--neutrino
spectrum. 

\subsection{Atmospheric data}

The atmospheric neutrino experiments measure $\nu_\mu$ and $\nu_e$ (and
their antineutrinos) created when cosmic rays interact with the Earth's
atmosphere. One expects roughly twice as many muon neutrinos as electron
neutrinos from the resulting cascade of pion and other meson decays.
Several experiments~\cite{atmos,SuperK} obtain a $\nu_\mu/\nu_e$ ratio
that is about 0.6 of the value expected from detailed theoretical
calculations of the flux~\cite{flux}. The Super-Kamiokande (SuperK)
experiment has collected the most data and analysis~\cite{SuperK}
indicates that their results for contained events can be explained as
$\nu_\mu\rightarrow\nu_\tau$ oscillations
with~\cite{SuperK,oldatmos,atmanal}
\begin{equation}
3\times10^{-4} {\rm~eV}^2 \le \delta m^2_{atm}
\le 7\times10^{-3} {\rm~eV}^2 \,, \quad
0.8 \le A^{\mu\not\mu}_{atm} \le 1.0 \;.
\label{atmdata}
\end{equation}
The high end of each range is favored.

Independent of flux normalization considerations, the
$\nu_\mu\rightarrow\nu_e$ oscillation channel is strongly disfavored by
the zenith angle distributions of the data~\cite{SuperK} and by the
up/down asymmetry separated into ``muon--like'' (\numu) and
``electron--like''(\nue) events~\cite{SKam:up/down}, which yield an
up--to--down ratio of $0.52^{+0.07}_{-0.06}\pm0.01$ for
$\mu$--like events and $0.84^{+0.14}_{-0.12}\pm0.02$ for $e$--like
events (the expected values are close to unity). Furthermore, the
recent CHOOZ $\bar\nu_e$ disappearance experiment excludes $\bar\nu_e
\rightarrow \bar\nu_\mu$ oscillations with large mixing
$A^{\mu\not\mu}_{atm} \gsim 0.2$
for $\delta m^2_{atm} \geq 10^{-3} {\rm eV}^2$~\cite{CHOOZ}.

In a four--neutrino context, another possibility for the atmospheric
neutrino oscillations is $\nu_\mu\rightarrow\nu_s$.  Oscillations of
this type in principle could be affected by matter due to the different
neutral current interactions of $\nu_\mu$ and $\nu_s$.  However, for the
contained events (with lower energy) these effects are small, especially
for larger values of $\delta m^2$ \cite{ls97,lipari}; hence, the allowed
regions for $\nu_\mu\rightarrow\nu_s$ should be similar to those for
$\nu_\mu\rightarrow\nu_\tau$. For events at higher energies the matter
effects could begin to be appreciable; a definitive test requires more
data.

\subsection{Solar data}

The solar neutrino experiments~\cite{solar} measure $\nu_e$ created in
the sun. There are three types of experiments, $\nu_e$ capture in Cl in
the Homestake mine, $\nu_e-e$ scattering at Kamiokande and
Super-Kamiokande, and $\nu_e$ capture in Ga at SAGE and GALLEX; each is
sensitive to different ranges of the solar neutrino spectrum and
measures a suppression from the expectations of the standard solar
model (SSM)\cite{SSM}.

For $\nu_e\rightarrow\nu_s$ oscillations in the sun (in which case
atmospheric neutrino oscillations are $\nu_\mu\rightarrow\nu_\tau$ in
our model) the allowed parameter ranges at 95\%~C.L.~\cite{hata} for the
small-angle matter-enhanced solution are given in
Table~\ref{sunsol}. The solution is based on the SSM fluxes in
Ref.~\cite{SSM}. Approximate parameters for the large-angle
matter-enhanced \cite{hata} and vacuum long--wavelength solutions
\cite{krastev2} for $\nu_e\rightarrow\nu_s$ oscillations of solar
neutrinos are also shown in Table~\ref{sunsol}. If the solar neutrino
deficit is caused instead by $\nu_e\rightarrow\nu_\tau$ oscillations
(and the atmospheric oscillations are $\nu_\mu\rightarrow\nu_s$), then
the allowed solar parameter ranges for the three solar cases are
slightly different \cite{hata}; see Table~\ref{sunsol}. The exact values
of the parameters may change as new data from SuperK \cite{totsuka}
become available and when fits are made with the new solar flux
calculations.

In any of the matter--enhanced scenarios it is also necessary that the
eigenmass $m_1$ associated predominantly with $\nu_e$ be lighter than
the eigenmass $m_0$ associated predominantly with the neutrino into
which the $\nu_e$ is oscillating (i.e., \nus\ or $\nu_\tau$), so that it
is $\nu_e$ rather than ${\bar\nu}_e$ that is resonant in the sun. For
the vacuum solutions the ordering of $m_0$ and $m_1$ does not matter.
Alternate scenarios where the $\nu_e$ is predominantly associated with
the heavier two states and $\nu_\mu$ is predominantly associated with
the lighter two states are also viable.

For $\nu_e\rightarrow\nu_\tau$ oscillations in the two-neutrino
approximation the propagation equation for
the neutrino states in the charge-current basis is~\cite{MSW1,bppw,lang}
\beq
i {d \over dt} \left( \begin{array}{c} \nu_e \\ \nu_\tau \end{array}
\right) = {1\over4E} \left(
\begin{array}{cc}
4\sqrt2 G_F E N_e & \delta m^2 \sin2\theta \\ 
\delta m^2 \sin2\theta & 2 \delta m^2 \cos2\theta
\end{array} \right)
\left( \begin{array}{c} \nu_e \\ \nu_\tau \end{array} \right) \,,
\eeq
where $N_e$ is the electron number density. For
$\nu_e\rightarrow\nu_s$ oscillations the propagation equation is
instead~\cite{MSWsterile}
\beq
i {d \over dt} \left( \begin{array}{c} \nu_e \\ \nu_s \end{array}
\right) = {1\over4E} \left(
\begin{array}{cc}
4\sqrt2 G_F E (N_e-{1\over2}N_n) & \delta m^2 \sin2\theta \\ 
\delta m^2 \sin2\theta & 2 \delta m^2 \cos2\theta
\end{array} \right)
\left( \begin{array}{c} \nu_e \\ \nu_s \end{array} \right) \,,
\eeq
where $N_n$ is the neutron number density.
For the small--angle matter--enhanced case the non--adiabatic approximate
solution for neutrino propagation is appropriate and the oscillation
probability for a neutrino of energy $E_\nu$ is
\begin{equation}
P(\nu_e\rightarrow\nu_e) = {1\over2}A^{e\not e}_{sun}
+P_x(1-A^{e\not e}_{sun}) \,
\end{equation}
where in this case $\not e$ labels either $\tau$ or $s$, and
\begin{equation}
P_x=exp\left[-{\pi\delta m^2_{sun} (A_{sun}^{e\not e})^2 \over
4 E_\nu \sqrt{1-A^{e\not e}_{sun}} \, (d\log N/dL)_c} \right] \,
\end{equation}
is the Landau--Zener transition probability and $N$ is either $N_e$ (for
$\nu_e\rightarrow\nu_\tau$ oscillations) or $N_e-{1\over2}N_n$ (for
$\nu_e\rightarrow\nu_s$ oscillations). The quantity $(d\log N/dL)_c$
is the appropriate logarithmic density gradient in the sun at
$N^{crit} = \delta m^2_{sun} \sqrt{1-A^{e\not e}_{sun}}/(2\sqrt2 G_F
E_\nu)$, the critical density where maximal oscillations (resonance)
occur. For the large--angle matter--enhanced case,
the neutrino propagation is adiabatic and
\begin{equation}
P(\nu_e\rightarrow\nu_e) = {1\over2} \left[
1 - \sqrt{1-A^{e\not e}_{sun}} \, \right] \,
\end{equation}
assuming the neutrinos are created where the electron density is well
above the critical density. For the vacuum long--wavelength solution
the oscillation probability is just given by the usual vacuum
expressions.

\subsection{Oscillation lengths and amplitudes summarized}

In neutrino oscillation descriptions of the solar, atmospheric, and LSND
data, a distinct oscillation wavelength and oscillation amplitude is
required for each of the three data sets.  Experimental uncertainties
allow for some latitude in these amplitudes and wavelengths, and for the
solar data, there are three isolated islands of viability in the $\delta
m^2$--amplitude plane~\cite{MSWconf}; see Fig.~3. The day--night
asymmetry measurement, found to be small in the recent SuperK data,
removed about half of the previously viable solar regions
\cite{MSWconf}.

The vacuum oscillation wavelength is linear in the neutrino
energy, allowing further possibilities that are summarized in
Table~\ref{amplength}. The chosen neutrino energies are typical for
solar, and reactor sources (5 MeV), pion facilities (100 MeV), for
contained (2 GeV), partially--contained (10 GeV), and throughgoing (100
GeV) neutrino events in underground detectors, and for astrophysical
sources (1 TeV).  Although full oscillation wavelengths are also listed
in Table~\ref{amplength}, oscillation effects may well be
measurable for a fraction of an oscillation wavelength or as an average
over many oscillation wavelengths.  Throughgoing and partially contained
atmospheric neutrinos may show nodes as a function of L/E. Further
possibilities arise when the matter effect of the earth is included in
the oscillation physics.  We consider earth--matter effects in Sec.~7.6.

\subsection{Inferences from data}

We consider first the probability rectangles for the atmospheric and
CHOOZ data.  The atmospheric data indicate $\delta m^2_{atm} \sim
5\times 10^{-3}$ and nearly maximal flavor--changing mixing of $\nu_\mu$
with $\nu_{\not e}$.  The present data do not distinguish between
$\nu_\tau$ or $\nu_s$ as the dominant state into which $\nu_\mu$ mixes.
The probability rectangles for the atmospheric scale are displayed in
Fig.~4a.  We label the two masses defining the atmospheric scale as
\nutwo\ and \nuthree, with $\delta m^2_{atm}=\delta m^2_{32}$.  Because
of the ``$\nu_\tau$--$\nu_s$'' ambiguity we show the union
$P_{\tau}+P_s$ rather than the partitions into $P_{\tau}$ and $P_s$.
For maximal \numu\--$\nu_{\not e}$ mixing, one must choose $P_{\mu
2}\sim P_{\mu 3}\sim P_{\tau 2}+P_{s2}\sim P_{\tau 3}+P_{s3}\sim1/2$.

Next we consider the pair of mass eigenstates whose mass--squared
difference is fixed by the solar scale. We provisionally investigate a
four--neutrino mass spectrum that consists of two pairs of nearly
degenerate neutrinos separated by the LSND scale $\delta m^2_{sbl}\sim
{\rm eV}^2$ (to explain the LSND result in terms of oscillations). We
argue in Sec.~4 that the data favor this spectrum over a spectrum with
one mass separated from three relatively degenerate masses.  We label
the second pair of mass states as \nuzero\ and \nuone, and define
$\delta m^2_{sun}=\delta m^2_{10}$.  Since $P_{\mu 2}$ and $P_{\mu 3}$
sum to near unity, $P_{\mu 0}$ and $P_{\mu 1}$ must be small.  Thus the
probability rectangles for the $\nu_0$ and $\nu_1$ states appear as
shown in Fig.~4a.  Accordingly, the LSND amplitude for
$\nu_{\mu}\rightarrow \nu_e$,
\beq
A^{\mu e}_{sbl}=-4\,
[U_{\mu 3} U_{e 3} + U_{\mu 2} U_{e 2}]\,
[U_{\mu 1} U_{e 1} +  U_{\mu 0} U_{e 0}]\,,
\label{lsnd1}
\eeq
is necessarily small.
We emphasize that the smallness of the LSND $\nu_e$--$\nu_\mu$
mixing is an inevitable
consequence of the large mixing of \numu\ to $\nu_{\not e}$ at
the atmospheric scale and the constraints of unitarity, independent of
particular model considerations including rearrangement of the
neutrino mass spectrum.

Four--neutrino unitarity may be used to rewrite eqn. (\ref{lsnd1}) as
\beq
A^{\mu e}_{sbl}
=4\, |U_{\mu 3} U_{e3} + U_{\mu 2} U_{e2}|^2
=4\, |U_{\mu 1} U_{e1} + U_{\mu 0} U_{e0}|^2.
\label{lsnd2}
\eeq
Written this way, it is clear that the LSND data is blind to the
partitioning of \nutau\ and \nus\ in the probability rectangles
of mass states \nuzero\ and \nuone.
This flavor ambiguity is shown in Fig.~4a.

With the identification $\delta m^2_{sun}= \delta m^2_{10}$,
we may use Eq.\ (\ref{theorem}) to write the solar \nue --disappearance
amplitude as
\beq
A^{e\not e}_{sun}=4\,P_{e0} P_{e1}.
\label{Asun}
\eeq
Because matter in the sun may exert a significant effect on
propagating neutrinos, the values of $P_{e0}$ and $P_{e1}$ for the
sun have some sensitivity to the state \nutau\ or $\nu_s$ into which
$\nu_e$ oscillates. However, the sensitivity of present data to this
difference is weak, and there is considerable
freedom in assigning \nutau\ or \nus\ or a linear combination thereof
as the mixing partner to $\nu_e$.
This $\nu_\tau$--$\nu_s$ ambiguity for \nue\ mixing at the solar scale
is complementary to the $\nu_s$--$\nu_\tau$ ambiguity
for \numu\ mixing at the atmospheric scale.
Potential measurements to resolve the
$\nu_s$--$\nu_\tau$ ambiguity at the solar scale will be
discussed in Sec.~7.

\subsection{The three solar solutions}

The $P_{e0}$ and $P_{e1}$ partitioning specifies whether
the solar model is a small--angle model or a large--angle model.
As can be inferred from Eq. (\ref{Asun}),
with nearly--equal partitioning of $P_{e0}$ and $P_{e1}$,
the mixing amplitude is near maximal (large angle).
With highly nonequal partitioning, i.e., $P_{e0} \ll P_{e1}$ or
$P_{e1} \ll P_{e0}$, the mixing amplitude is small.
Of the three viable solar neutrino options, SAM falls into the small angle
category, while LAM and VLW fall into the large angle category.
The probability rectangles for the small and large angle classes of models
are shown in Figs.~4b and 4c.
Recall that in order to obtain the MSW resonant enhancement required
for the SAM and LAM solutions,
it is necessary that the state which is predominantly \nue\
be the lighter of the two mass states, \nuone.
Qualitatively, LAM and VLW are distinguishable in their probability
rectangles only by the choice of value for $\delta m^2_{sun}$.
Quantitatively, the two solutions and the VLW solution are distinguishable
in ways which are discussed in Sec.~7.

If the active--sterile mixing is small,
then all ambiguities in the probability rectangles are resolved:
the large atmospheric mixing must be
$\nu_\mu$--$\nu_\tau$, and the solar solution must be
small--angle SAM with $\nu_e$--$\nu_s$ mixing.
The probability rectangles for this model are shown in Fig.~2.
This particular solution  has recently been analyzed in the context of the
minimal four--neutrino mass matrix~\cite{bww98,gmnr98}.

\section{Mass spectra}

\subsection{Argument against a 1+3 mass spectrum}

It has been shown by Bilenky, Giunti and Grimus~\cite{bgg98} that a
hierarchical ordering of the four--neutrino spectrum (implying one
dominant mass) is disfavored by the data when the null results of
reactor and accelerator disappearance experiments are included. We will
refer to this spectrum as the 1+3 spectrum, defined
as one heavier mass state separated from three lighter,
nearly-degenerate states, or vice versa. We demonstrate the argument
with a set of logical steps similar to theirs.

Assume a mass spectrum with one heavy mass well separated from three other
nearly-degenerate states and let the heavy mass state be labeled
as $\nu_3$. Then the LSND mass-squared scale is $\delta m^2_{sbl}
\simeq \delta m^2_{32} \simeq \delta m^2_{31} \simeq \delta m^2_{30}$
and the LSND amplitude is
\beq
A^{\mu e}_{sbl}=-4\,U_{e3}U_{\mu 3}\, [\sum_{j\neq 3} U_{\mu j}U_{ej}]
                =4\,P_{e3} P_{\mu 3}, \qquad
{\rm with~} P_{e3}+P_{\mu 3}\le 1.
\label{emu1}
\eeq
On the other hand, the \nue\ and \numu\ disappearance experiments at
reactors and accelerators are also sensitive to the LSND scale.
These experiments measure the disappearance amplitudes
\beq
A^{e\not e}_{sbl}=4 \, P_{e3} \, [\sum_{j\neq 3} P_{e j}]
                   =4\,P_{e3} \, [1-P_{e3}]
\label{enote}
\eeq
and
\beq
A^{\mu\not\mu}_{sbl} = 4\,P_{\mu 3} \,[1-P_{\mu 3}].
\label{munotmu}
\eeq
The second equalities in Eqs.\ (\ref{emu1}) and (\ref{enote}) (see
Eq.~(\ref{theorem})) follow from unitarity of the mixing matrix.
The three amplitudes in Eqs.\ (\ref{emu1})--(\ref{munotmu})
depend on just two parameters, and so are interrelated.
All three of these amplitudes are constrained by experiments to be small.
{\it A priori} then, $P_{e3}$ and $P_{\mu 3}$ may both be small,
or one (but not both) may be near unity  with the other small.
The fact that $A^{\mu e}_{sbl}$ is an {\it appearance observation}
rather than a bound means that if $P_{e3}$ and $P_{\mu 3}$ are both small,
they cannot be too small.

In the 1+3 model, the atmospheric scale does not involve the heavy state
$\nu_3$. Without loss of generality we label the state which determines
the atmospheric scale as $\nu_2$. Then from Eq.~(\ref{theorem})
the atmospheric $\nu_\mu$ disappearance oscillation amplitude is given
by
\beq
A^{\mu\not\mu}_{atm} = 4 P_{\mu 2} (P_{\mu 0} + P_{\mu 1})
\leq (1-P_{\mu 3})^2 \,,
\label{atmbound}
\eeq
where the inequality comes from maximizing
$4P_{\mu 2} (P_{\mu 0} + P_{\mu 1})$ subject to the constraint
$P_{\mu 0} + P_{\mu 1} + P_{\mu 2} = 1 - P_{\mu 3}$.
The SuperK data indicate that \numu\ is maximally mixed at the
$\delta m^2_{atm}$ scale, i.e., there is little \numu--content
available to the $\nu_3$ state. Quantitatively we have
$A^{\mu\not\mu}_{atm} \geq 0.8$, which implies
$P_{\mu 3} \leq 0.11$. Since $P_{\mu 3}$ is small, Eq.~(\ref{munotmu})
becomes
\beq
A^{\mu\not\mu}_{sbl}\simeq 4\,P_{\mu 3} << 1.
\label{munotmu2}
\eeq
The probability rectangles for the 1+3
model with small $P_{\mu 3}$ are presented in Fig.~5.
Note that it is the zenith--angle, or up/down asymmetry data,
which really establishes $\delta m^2_{atm}$ as different from
$\delta m^2_{sbl}$, that is crucial for the argument~\cite{bgg98}.

We are left with the possibilities of $P_{e3}$ being small or near
unity. As can be seen in Fig.~5, if $P_{e3}$ is near unity, then there
is little $P_{e\not 3}$ to distribute over the three lighter mass
states. In particular, the solar amplitude $A^{e\not
e}_{sun}=4P_{e0}P_{e1}$, where the solar scale is
$\delta m^2_{sun}=\delta m^2_{10}$, is second order in small quantities,
too small for even the SAM solution ($A_{SAM}\ge 2.5\times 10^{-3}$)
to the solar flux. This may be easily quantified.
If $P_{e3}$ were near unity, we would have
\beq
A^{e\not e}_{sbl}\simeq 4\,[1-P_{e 3}] << 1.
\label{enote2}
\eeq
Together with unitarity, this in turn bounds the magnitude of the solar
amplitude:
\beq
A^{e\not e}_{sun}=4\,P_{e0} P_{e1}
   \, \leq \, (1-P_{e3})^2\simeq \frac{1}{16} (A^{e\not e}_{sbl})^2,
\label{sunbound}
\eeq
where the inequality in Eq.~(\ref{sunbound}) comes from maximizing
$4P_{e0}P_{e1}$ subject to the constraint $P_{e0}+P_{e1} \leq 1-P_{e3}$.
The experimental upper limit on $A^{e\not e}_{sbl}$ from the BUGEY
experiment~\cite{Bugey} is about 0.1 for $\delta m^2_{sbl}
\leq 2$~eV$^2$, which disallows even the small--angle solar
solution. We conclude that $P_{e3}$ is small, in which case
\beq
A^{e\not e}_{sbl} \simeq 4\,P_{e3} << 1.
\label{enote3}
\eeq

Thus, both $P_{e3}$ and $P_{\mu 3}$ must be small in the 1+3 model,
and from Eqs.~(\ref{emu1}), (\ref{munotmu2}), and (\ref{enote3}),
we infer the relation
\beq
A^{\mu e}_{sbl}\simeq \frac{1}{4}A^{\mu\not\mu}_{sbl} \; A^{e\not e}_{sbl}.
\label{AAA}
\eeq
However, the experimental upper bounds on the disappearance amplitudes
$A^{e\not e}_{sbl}$ \cite{Bugey} and $A^{\mu\not\mu}_{sbl}$
\cite{CDHS} and the measured appearance result for $A^{\mu e}_{sbl}$
\cite{LSND} are not compatible with Eq.~(\ref{AAA}), thereby
disfavoring the 1+3 model. For example, for $\delta m^2_{sbl} =
0.3$~eV$^2$, $A^{e\not e}_{sbl} < 0.035$ from Bugey,
$A^{\mu\not\mu}_{sbl} < 0.8$ from CDHS, which implies
$A^{\mu e}_{sbl} < 0.007$; however, for this value of $\delta
m^2_{sbl}$, the LSND data indicate $A^{\mu e}_{sbl} > 0.04$. The LSND
results are presented in terms of maximum likelihood rather than
confidence level limits, so it is not straightforward to state an
exclusion probability.

Put another way, $A^{\mu e}_{sbl}$ is large enough that the Bugey and
CDHS limits force one of $P_{e 3}$ and $P_{\mu 3}$ to be small and the
other to be large, but this is ruled out by the solar and atmospheric
data. The constraints on $P_{e3}$ and $P_{\mu 3}$
from the three short--baseline amplitudes $A^{e \mu}_{sbl}$,
$A^{e\not e}_{sbl}$, and $A^{\mu\not\mu}_{sbl}$, the atmospheric
amplitude $A^{\mu\not\mu}_{atm}$ and the solar amplitude $A^{e\not
e}_{sun}$ (from Eqs.~(\ref{emu1}), (\ref{enote}), (\ref{munotmu}),
(\ref{atmbound}), and (\ref{sunbound}), respectively) are conveniently
summarized in Fig.~6 for two different values of $\delta m^2_{sbl}$.

The measured values and bounds for the short--baseline appearance and
disappearance amplitudes depend on the magnitude of $\delta m^2_{sbl}$.
(There is effectively a suppressed $\delta m^2_{sbl}$ third axis
in our Fig.~6, which samples only two particular values of $\delta
m^2_{sbl}$.) For certain allowed values of $\delta m^2$
(e.g., at 1.7 ${\rm eV}^2$ and 0.25 ${\rm eV}^2$ according to
Fig.~2 of \cite{bgg98}) the violation of Eq.~(\ref{AAA}) is mild,
and the 1+3 model is just barely incompatible with the data; see,
e.g., Fig.~6a.

The argument against the 1+3 model does not depend on the sign of
$\delta m^2_{sbl}$. This means that the inverted 3+1 model with the
three nearly degenerate mass states heavier than the remaining state is
equally disfavored.

\subsection{2+2 mass spectrum}

We now turn to the favored class of four--neutrino models,
namely those with two nearly degenerate mass pairs
separated by the LSND scale as displayed in Fig.~1.
It is interesting to see how this ``pair of pairs'' mass spectrum of
four--neutrino models realizes the dependency  among
$A^{\mu e}_{sbl}$, $A^{\mu\not\mu}_{sbl}$, and $A^{e\not e}_{sbl}$
which conflicted with the 1+3 model. Let $\nu_0$ and $\nu_1$ label
the pair of the nearly--degenerate mass eigenstates responsible for the
solar oscillations, and $\nu_2$ and $\nu_3$ label
the pair of the nearly--degenerate mass eigenstates responsible for the
atmospheric oscillations.

The expressions for the oscillation amplitudes are
\beq
A^{\mu e}_{sbl}=
4\,|U_{e2}U_{\mu 2}+U_{e3}U_{\mu 3}|^2 =
4\,|U_{e0}U_{\mu 0}+U_{e1}U_{\mu 1}|^2,
\label{emupairs}
\eeq
\beq
A^{e\not e}_{sbl}
= 4(P_{e3}P_{e1}+P_{e3}P_{e0}+P_{e2}P_{e1}+P_{e2}P_{e0})
= 4\,\sigma_e\,(1-\sigma_e),
\label{enotepairs}
\eeq
and
\beq
A^{\mu\not\mu}_{sbl} = 4(P_{\mu 3}P_{\mu1} + P_{\mu 3}P_{\mu 0}
+ P_{\mu 2}P_{\mu1} + P_{\mu2}P_{\mu0})
= 4\,\sigma_\mu\,(1-\sigma_\mu),
\label{munotmupairs}
\eeq
with
\beq
\sigma_\alpha \equiv |U_{\alpha 2}|^2+|U_{\alpha 3}|^2
= P_{\alpha 2} + P_{\alpha 3} \,.
\label{sigma}
\eeq
The Schwartz vector inequality
$|\vec{v}_e\cdot\vec{v}_\mu|^2 \leq |\vec{v}_e|^2\,|\vec{v}_\mu|^2$
applied to the vectors $\vec{v}_e\equiv(U_{e2},U_{e3})$ and
$\vec{v}_\mu\equiv(U_{\mu 2},U_{\mu 3})$ then gives
\beq
A^{\mu e}_{sbl} \leq 4\,\sigma_e\,\sigma_\mu \,.
\label{schwartz}
\eeq
Furthermore, in the 2+2 model the solar oscillation amplitude is
\beq
A^{e\not e}_{sun} = 4 P_{e0} P_{e1} \leq (1 - \sigma_e)^2 \,,
\label{sunboundpairs}
\eeq
and the atmospheric oscillation amplitude is
\beq
A^{\mu\not\mu}_{atm} = 4 P_{\mu 2} P_{\mu 3} \leq \sigma_\mu^2 \,,
\label{atmboundpairs}
\eeq
where the inequalities in Eqs.~(\ref{sunboundpairs}) and
(\ref{atmboundpairs}) come from maximizing the expressions subject
to the constraints $P_{e0} + P_{e1} = 1 - \sigma_e$ and $P_{\mu 2}
+ P_{\mu 3} = \sigma_\mu$, respectively.

If the vector inequality in Eq.~(\ref{schwartz}) is saturated, then
$A^{e\mu}_{sbl}$, $A^{e\not e}_{sbl}$, and $A^{\mu\not\mu}_{sbl}$
each has the same functional dependence on two parameters as it did
in the 1+3 model ($\sigma_e$ has replaced $P_{e3}$ and $\sigma_\mu$
has replaced $P_{\mu 3}$). Then the previous argument that the LSND,
Bugey and CDHS data require one parameter to be small ($\ll 1$) and
the other large ($\simeq 1$) applies. The argument is unaffected if
the vector inequality is not saturated. As before in the 1+3 case, the
solar constraint indicates that $\sigma_e$ must be small. This time
however, unlike the 1+3 case, the atmospheric constraint involves
$\sigma_\mu^2$ and not $(1-\sigma_\mu)^2$, and can be met if
$\sigma_\mu$ is large ($\simeq1$). Therefore the constraints of the
data can be satisfied by assigning $\nu_e$ dominantly to one pair of
mass states and $\nu_\mu$ dominantly to the other pair. Instead of
Eq.~(\ref{AAA}) pertinent to the 1+3 spectrum, we obtain for the 2+2
spectrum
\beq
A^{\mu e}_{sbl} \le A^{e\not e}_{sbl} \,.
\label{AA}
\eeq
This bound is linear in the small disappearance amplitudes, and is easily
satisfied by the data. For example, the tightest constraint on
$A^{e\not e}_{sbl}$ is about 0.02 for $\delta m^2_{sbl}=0.6$~eV$^2$,
while the LSND data indicate $A^{\mu e}_{sbl}$ can be as low as 0.009
for this value of $\delta m^2_{sbl}$.

In Fig.~7 we have drawn the $\sigma_e$--$\sigma_\mu$ plot for the 2+2
model, analogous to the $P_{e3}$--$P_{\mu 3}$ plot for the 1+3 model,
for $\delta m^2_{sbl}=1.7$~eV$^2$.
The allowed regions with $\sigma_\mu$ near unity (implying
near--maximal mixing of $\nu_\mu$ in the $\nu_2$--$\nu_3$ pair)
and $\sigma_e$ small (implying almost no mixing of $\nu_e$ into
the $\nu_2$--$\nu_3$ pair) show that the 2+2 model can
comfortably accommodate the data.

Since only mass-squared differences are important for oscillations, the
inverted 2+2 model, where the solar oscillations are driven by the
mass-squared difference of the upper mass pair and the atmospheric
oscillations are driven by the mass-squared difference of the lower mass
pair, is equally viable.

\subsection{New results}


Two features of the data are especially noteworthy.
The first is the remarkably high degree of isolation of
$\nu_e$ into one mass pair and $\nu_\mu$ into the other mass pair,
as inferred from the bounds on the disappearance amplitudes.
The second is the near saturation of the vector inequality in
Eq.~(\ref{schwartz}) by the LSND appearance amplitude
$A^{\mu e}_{sbl}$ for $\delta m^2_{sbl}\simeq0.3$~eV$^2$.

Equations~(\ref{enotepairs}) and (\ref{munotmupairs}) bound
the degree to which \nue\ and \numu\ are found in opposite pairs
of mass eigenstates. Without loss of generality we assume that $\nu_e$
is predominantly associated with $\nu_0$ and $\nu_1$, and that
$\nu_\mu$ is predominantly associated with $\nu_2$ and $\nu_3$. Then
from the Bugey and CDHS data we find the constraints
\beq
\sigma_e \simeq \frac{1}{4} \, A^{e\not e}_{sbl} \le 0.016 \, (0.009) \,,
\label{sigesmall}
\eeq
and
\beq
1-\sigma_\mu \simeq \frac{1}{4}\, A^{\mu\not\mu}_{sbl} \le 0.013 \, (0.2) \,,
\label{sigmusmall}
\eeq
respectively, for $\delta m^2_{sbl}=2$~eV$^2$ ($0.3$~eV$^2$). We then
deduce
\beq
A^{e\not e}_{sbl} \leq 0.065 \, (0.04) \,,
\eeq
which can be compared to the LSND data
\beq
A^{\mu e}_{sbl} \simeq 0.0025 \, (0.04) \,,
\eeq
for these two values of $\delta m^2_{sbl}$. 
The near--saturation of the inequality in Eq.~(\ref{AA}) for
$\delta m^2_{sbl}=0.3$~eV$^2$ has very interesting implications.
It means that $\hat{v}_e$ is nearly parallel or antiparallel
to $\hat{v}_\mu$, which in turn indicates that
\beq
|U_{e2}/U_{e3}| \simeq |U_{\mu 2}/U_{\mu 3}| \,.
\label{parallel23}
\eeq
%
%
This is a new result.

Furthermore, the SuperK data suggest that \numu\ is maximally
mixed in the mass pair with mass-squared difference 
$\delta m^2_{atm}$, so for this pair, called $\nu_2$ and $\nu_3$, that
\beq
|U_{\mu 2}| \simeq |U_{\mu 3}| \simeq {1\over\sqrt2}.
\label{Umu23}
\eeq
Then Eq.~(\ref{parallel23}) implies
\beq
|U_{e2}| \simeq |U_{e3}|.
\label{Ue23}
\eeq
This is also a new result. In summary, if the oscillation parameters are
indeed near the limits of the Bugey bound, the four--neutrino mixing
matrix in the 2+2 model must satisfy Eq.~(\ref{parallel23}), which
implies Eq.~(\ref{Ue23}) if the atmospheric $\nu_\mu$ mixing is maximal.

We can derive additional constraints by considering $\sigma_e^\prime =
P_{e0} + P_{e1}$ and $\sigma_\mu^\prime = P_{\mu 0} + P_{\mu 1}$ rather
than $\sigma_e$ and $\sigma_\mu$. The
data requires $\sigma_e^\prime$ to be large ($\simeq 1$) and
$\sigma_\mu^\prime$ small, and the Schwartz inequality reduces to
\beq
A^{\mu e}_{sbl} \leq A^{\mu\not\mu}_{sbl} \,,
\label{schwartz2}
\eeq
where the CDHS bound is
\beq
A^{\mu\not \mu}_{sbl} \leq 0.05 \, (0.8) \,,
\eeq
at $\delta m^2_{sbl}=2.0$~eV$^2$ ($0.3$~eV$^2$).
Because the inequality in Eq.~(\ref{schwartz2}) is not saturated by the
data for any $\delta m^2_{sbl}$, a relation similar to
Eq.~(\ref{parallel23}) for $U_{e0}$, $U_{e1}$, $U_{\mu 0}$,
and $U_{\mu 1}$ is not required in the 2+2 model. However, the explicit
mass matrices we consider do have such additional relations; see Sec.~5.

Finally, we mention a curiosity~\cite{raffelt} in the data
which occurs for the pair of pairs mass spectrum with the matter--enhanced
solar solutions (SAM and LAM).  The {\it linear} mass splitting at the
heavier pair is $m_3-m_2 \sim \delta m_{32}^2/2m_3$.  If this pair is
associated with the atmospheric scale, we have $m_3-m_2 \sim 2.5\times 10^{-2}
{\rm eV}\,(\delta m^2_{atm}/5\times 10^{-3}{\rm eV}^2)(m_3/{\rm eV})^{-1}$.
On the other hand, if the lighter mass pair is associated with the
matter--enhanced solar scale, and $m_0 \gg m_1$,
then the {linear} mass--splitting of this pair is
$m_0-m_1 \sim m_0 \sim \sqrt{\delta m^2_{sun}}=
2.5\times 10^{-2}\,{\rm eV}\,
(\delta m^2_{sun}/10^{-5}{\rm eV}^2)^\frac{1}{2}$.
The two linear mass splittings within the pairs are nearly identical.
While squared masses enter into the oscillation formulae for
relativistic neutrinos, the more fundamental constructs of field theory,
such as the  Lagrangian and the
resulting equations of motion, are linear in fermion masses
(and quadratic in boson masses, these powers of mass being
related to the dimensionality of the fermion field vs. the boson field).
Thus it is a worthy enterprise to attempt to deduce linear neutrino--mass
relations whenever possible.

\section{Mass matrix ansatzes}

\subsection{Solar $\nu_e\rightarrow\nu_s$ oscillations}

To describe the above oscillation
phenomena in the scenario where the solar neutrino deficit is described
by $\nu_e\rightarrow\nu_s$ oscillations and the atmospheric data by
$\nu_\mu\rightarrow\nu_\tau$, we consider the neutrino mass matrix ansatz
\begin{equation}
M = m \left( \begin{array}{cccc}
\eps_1 & \eps_2 & 0      & 0\\
\eps_2 & 0      & 0      & \eps_3\\
0      & 0      & \eps_4 & 1\\
0      & \eps_3 & 1      & \eps_4
\end{array} \right) \;,
\label{m}
\end{equation}
presented in the ($\nu_s, \nu_e, \nu_{\mu}, \nu_{\tau}$) basis
(i.e. the basis where the charged lepton mass matrix is diagonal). 
By considering the field redefinitions $\Psi \rightarrow -\Psi$ and
$\Psi \rightarrow \gamma_5\Psi$ one realizes that $m$, and at least one
of $\eps_1$ and $\eps_4$, and at least one of $\eps_2$ and $\eps_3$, may
be taken as positive; we will take $m$ and all four $\eps_j$ to be
positive for simplicity. The mass matrix $M$ contains five parameters
($m$, $\eps_1$, $\eps_2$, $\eps_3$, $\eps_4$), just enough to
incorporate the required three mass-squared
differences and the oscillation amplitudes for solar and LSND neutrinos.
The large amplitude for atmospheric oscillations does not require a
sixth parameter in our model because the structure of the mass matrix
naturally gives maximal mixing of $\nu_\mu$ with $\nu_\tau$ (or with
$\nu_s$ if $\nu_\tau$ and $\nu_s$ are interchanged).

For simplicity, we have taken the mass matrix to be real and symmetric.
The choice of a symmetric neutrino mass matrix is well--motivated in the
context of oscillations, for what is measured in neutrino oscillations
are the differences of squared masses, which are eigenvalues of the
hermitian matrix $M M^{\dag}$, which is itself symmetric when CP
conservation is assumed. $M$ is diagonalized by an orthogonal matrix $U$
(real) and there is no CP violation. The $\eps_j$ are assumed to be
small compared to unity, but not all necessarily of the same order of
magnitude. The zero terms in the mass matrix could be taken as nonzero
without changing the phenomenology discussed here as long as they are
small compared to the terms shown.  Also, the $M_{\nu_\mu \nu_\mu}$ term
could be chosen different from $\eps_4$ while still giving maximal
mixing of \numu\ and \nutau\ since maximal mixing results from the large
value of the $M_{\nu_\mu \nu_\tau}$ matrix element relative to the
diagonal $M_{\nu_\mu \nu_\mu}$ and $M_{\nu_\tau \nu_\tau}$ elements,
without any need for fine tuning of the difference $|M_{\nu_\mu
\nu_\mu}-M_{\nu_\tau \nu_\tau}|$.  Here we choose to take the minimal
form for $M$ needed to describe the data and then derive the associated
consequences.

To a good approximation, the two large eigenvalues of the
mass matrix in Eq.~(\ref{m}) are
\begin{equation}
\ts m_{2,3} = \mp m \left( 1 \mp \eps_4 +
{1\over2}\eps_3^2\right)\;.
\label{m23}
\end{equation}
The values of the two small eigenvalues depend on the hierarchy of the
$\epsilon_j$. For the three solar cases we have:
\begin{eqnarray}
{\rm SAM}: & \eps_2 \ll \eps_1, \eps_4 \ll \eps_3
\ll 1 \, \,,
\label{ordera}\\
{\rm LAM}: & \eps_1, \eps_2, \eps_4 \ll \eps_3 \ll
1 \, \,,
\label{orderb}\\
{\rm VLW}: & \eps_1 \ll \eps_2 \ll \eps_4 \ll \eps_3 \ll
1 \,.
\label{orderc}
\end{eqnarray}
The two small eigenvalues are then approximately given by
\begin{eqnarray}
{\rm SAM}: & m_0 \simeq m\eps_1\,, \quad
m_1 \simeq m(\eps_3^2 \eps_4 - \eps_2^2/\eps_1) \, \,,
\label{casea}\\
{\rm LAM}: & m_{0,1} \simeq {m\over2}
\left[\eps_1 \pm \sqrt{\eps_1^2+4\eps_2^2}\right] \, \,,
\label{caseb}\\
{\rm VLW}: & m_{0,1} \simeq m
\left[\pm \eps_2 + \eps_3^2\eps_4/2 \right] \,.
\label{casec}
\end{eqnarray}
These approximate expressions for the eigenvalues have been obtained by
multiplying each $\eps_j$ by powers of a hypothetical
parameter $\delta$, where the number of powers of $\delta$ assigned to
each $\eps_j$ depends upon the ordering in
Eqs.~(\ref{ordera})-(\ref{orderc}). For example, in the SAM case
$\eps_3$ are multiplied by $\delta$, $\eps_1$ and
$\eps_4$ by $\delta^2$, and $\eps_2$ by $\delta^3$. Then
each eigenvalue is written as an expansion in powers of $\delta$, the
coefficients of which may be solved for by requiring that the expression
$\Pi_i(\lambda-\lambda_i)$ reproduces the eigenvalue equation for the
mass matrix order by order in $\delta$. Once the coefficients are found,
$\delta$ is set equal to unity.

The eigenvalues in all cases have the desired hierarchy
$m_1 < m_0 \ll m_2, m_3$, which gives the mass spectrum of the 2+2 model
described in Sec.~4.2 and depicted in Fig.~1. The small relative mass
splitting of the heavier masses $m_2,m_3$ is governed entirely by the
parameter $\eps_4$: $\delta m^2_{32} \simeq 4 m^2 \eps_4$. The LSND
$\nu_\mu \rightarrow \nu_e$ oscillations are driven by the scale
$\delta m^2_{21} \simeq \delta m^2_{31} \simeq \delta m^2_{20} \simeq
\delta m^2_{30} \simeq m^2$, the atmospheric $\nu_\mu$ oscillations
are determined by $\delta m^2_{32}$, and the
solar $\nu_e \rightarrow \nu_s$ oscillations are determined by $\delta
m^2_{10}$, the approximate expression for which can be obtained by
Eqs.~(\ref{casea})-(\ref{casec}). The charged-current
eigenstates are approximately related to the mass eigenstates by
\begin{equation}
\left( \begin{array}{c}
\nu_s \\ \nu_e \\ \nu_\mu \\ \nu_\tau
\end{array} \right) = U
\left( \begin{array}{c}
\nu_0 \\ \nu_1 \\ \nu_2 \\ \nu_3
\end{array} \right) \simeq
\left( \begin{array}{cccc}
\cos\theta             & -\sin\theta
& \cdots                    & \cdots \\
\sin\theta             & \cos\theta
& {1\over\sqrt2}\eps_3 & {1\over\sqrt2}\eps_3 \\
-\eps_3\sin\theta      & -\eps_3\cos\theta
& {1\over\sqrt2}       & {1\over\sqrt2} \\
\cdots                      & \cdots
& -{1\over\sqrt2}      & {1\over\sqrt2}
\end{array} \right)
\left( \begin{array}{c}
\nu_0 \\ \nu_1 \\ \nu_2 \\ \nu_3
\end{array} \right) \;,
\label{U}
\end{equation}
where $\tan 2\theta = 2\eps_2/\eps_1$.  The dots indicate nonzero
terms that are much smaller than the terms shown.  It is their smallness
that suppresses mixing between \nutau\ and \nus.  The mixing matrix
$U$ depends on just three of the original five parameters; it is
independent of $\eps_4$ and the overall
mass--scale parameter $m$. Note that $\nu_0$ and $\nu_1$ couple
predominantly to $\nu_s$ and $\nu_e$.  The nearly--degenerate $\nu_2$
and $\nu_3$ are seen to consist primarily of nearly equal mixtures of
$\nu_\mu$ and $\nu_\tau$. These results, illustrated in Fig.~2, conform
to the qualitative arguments of Sec.~3 based on probability
rectangles.

It is noted that this mixing matrix not only satisfies the approximate
equalities of Eqs.~(\ref{parallel23})--(\ref{Ue23}), but in fact
replaces the
approximate equalities, derived from parameter--independent
arguments, with exact equalities to first order in $\eps_j$.
Inspection of the mixing matrix reveals that our model predicts
saturation of Eqs.~(\ref{AA}) and (\ref{schwartz2}) to this order, i.e.,
$A^{e\not e}_{sbl} = A^{\mu\not\mu}_{sbl} = A^{\mu e}_{sbl}$.
A small improvement in the measurement of $A^{e\not e}_{sbl}$
or a modest improvement in the measurement of $A^{\mu\not\mu}_{sbl}$
is predicted to show a positive disappearance signal.

\subsection{Solar $\nu_e\rightarrow\nu_\tau$ oscillations}

Another scenario, with solar $\nu_e\rightarrow\nu_\tau$ and
atmospheric $\nu_\mu\rightarrow\nu_s$ oscillations, is readily obtained
by interchanging $\nu_\tau \rightarrow \nu_s$ and $\nu_s \rightarrow
-\nu_\tau$. The mass matrix in the ($\nu_s$, $\nu_e$, $\nu_\mu$,
$\nu_\tau$) basis is then
\begin{equation}
M = m \left( \begin{array}{cccc}
\eps_4 &  \eps_3 & 1      & 0\\
\eps_3 &  0      & 0      & -\eps_2\\
1      &  0      & \eps_4 & 0\\
0      & -\eps_2 & 0      & \eps_1
\end{array} \right) \;.
\label{m2}
\end{equation}
The eigenvalues and parameter hierarchies are still given by
Eqs.~(\ref{m23})-(\ref{casec}). The mixing matrix is then given by
\begin{equation}
\left( \begin{array}{c}
\nu_s \\ \nu_e \\ \nu_\mu \\ \nu_\tau
\end{array} \right) = U
\left( \begin{array}{c}
\nu_0 \\ \nu_1 \\ \nu_2 \\ \nu_3
\end{array} \right) \simeq
\left( \begin{array}{cccc}
\cdots                       & \cdots
& -{1\over\sqrt2}            & {1\over\sqrt2} \\
\sin\theta                   & \cos\theta
& {1\over\sqrt2}\eps_3       & {1\over\sqrt2}\eps_3 \\
-\eps_3\sin\theta            & -\eps_3\cos\theta
& {1\over\sqrt2}             & {1\over\sqrt2} \\
-\cos\theta                  & \sin\theta
& \cdots                     & \cdots
\end{array} \right)
\left( \begin{array}{c}
\nu_0 \\ \nu_1 \\ \nu_2 \\ \nu_3
\end{array} \right) \;,
\label{U2}
\end{equation}
where again $\tan2\theta=2\eps_2/\eps_1$.

In the VLW case, the parameter $\eps_1$ is negligibly small if the
solar oscillations are maximal, and can be taken as zero without
affecting the phenomenology.  If this is done, then reference to the
mass matrix shows that both $\nu_e$ and $\nu_\tau$ derive their masses
entirely from flavor non--diagonal couplings, and they are maximally
mixed (analogous to the $\nu_\mu$--$\nu_s$ system).  Also, if
$\eps_1$ is taken as zero, then there are only four independent
parameters needed in the mass matrix, and just two in the mixing matrix.
The derived $\theta$ parameter becomes $\pm\frac{\pi}{4}$, and the
mixing matrix becomes very simple:
\begin{equation}
U \simeq \frac{1}{\sqrt{2}}
\left( \begin{array}{cccc}
\cdots        & \cdots     & -1           & 1 \\
\pm 1         &  1         & \eps_3       & \eps_3 \\
\mp \eps_3    & -\eps_3    & 1            & 1 \\
-1            & \pm 1      & \cdots       & \cdots
\end{array} \right)
\label{UVLW}
\end{equation}

\subsection{Solar $\nu_e\rightarrow\nu_s$ and
$\nu_e\rightarrow\nu_\tau$ oscillations}

A more general scenario which is a mixture of the previous two is for
solar neutrinos to undergo both $\nu_e\rightarrow\nu_s$ and
$\nu_e\rightarrow\nu_\tau$ oscillations. This is easily parameterized
by replacing the $\nu_s$ and $\nu_\tau$ states in Eq.\ (\ref{U}) with
the rotated states $\nu_s^\prime$ and $\nu_\tau^\prime$ and defining
\beq
\left( \begin{array}{c}
\nu_s^\prime \\ \nu_\tau^\prime
\end{array} \right) =
\left( \begin{array}{cc}
\cos\alpha	& -\sin\alpha \\
\sin\alpha	& \cos\alpha
\end{array} \right)
\left( \begin{array}{c}
\nu_s \\ \nu_\tau
\end{array} \right) \;.
\label{Urot}
\eeq
Then the mass matrix in the ($\nu_s$, $\nu_e$, $\nu_\mu$,
$\nu_\tau$) basis becomes
\begin{equation}
M = m \left( \begin{array}{cccc}
\eps_1\cos^2\alpha+\eps_4\sin^2\alpha
& \eps_2\cos\alpha+\eps_3\sin\alpha & \sin\alpha
& (\eps_4-\eps_1)\sin\alpha\cos\alpha \\
\eps_2\cos\alpha+\eps_3\sin\alpha & 0      & 0
& \eps_3\cos\alpha -\eps_2\sin\alpha\\
\sin\alpha      & 0      & \eps_4 & \cos\alpha \\
(\eps_4-\eps_1)\sin\alpha\cos\alpha
& \eps_3\cos\alpha -\eps_2\sin\alpha
& \cos\alpha & \eps_4\cos^2\alpha+\eps_1\sin^2\alpha
\end{array} \right) \;,
\label{mprime}
\end{equation}
and the matrix which diagonalizes $M$ is
\begin{equation}
\left( \begin{array}{c}
\nu_s \\ \nu_e \\ \nu_\mu \\ \nu_\tau
\end{array} \right) = U
\left( \begin{array}{c}
\nu_0 \\ \nu_1 \\ \nu_2 \\ \nu_3
\end{array} \right) =
\left( \begin{array}{cccc}
\cos\theta\cos\alpha          & -\sin\theta\cos\alpha
& -{1\over\sqrt2}\sin\alpha   & {1\over\sqrt2}\sin\alpha \\
\sin\theta                    & \cos\theta
& {1\over\sqrt2}\eps_3  & {1\over\sqrt2}\eps_3 \\
-\eps_3\sin\theta          & -\eps_3\cos\theta
& {1\over\sqrt2}              & {1\over\sqrt2} \\
-\cos\theta\sin\alpha         & \sin\theta\sin\alpha
& -{1\over\sqrt2}\cos\alpha   & {1\over\sqrt2}\cos\alpha
\end{array} \right)
\left( \begin{array}{c}
\nu_0 \\ \nu_1 \\ \nu_2 \\ \nu_3
\end{array} \right) \;,
\label{Uprime}
\end{equation}
where $\tan2\theta=2\eps_2/\eps_1$ as before.

\section{Oscillation probabilities}

\subsection{Expressions for any baseline}

For the mixing in
Eq.\ (\ref{U}) (when the solar oscillations are $\nu_e\rightarrow\nu_s$),
the off-diagonal vacuum oscillation probabilities obtained from
Eq.\ (\ref{oscprob}), to
leading order in $\eps_j$ for each $\Delta_{ij}$ and ignoring
amplitudes smaller than ${\cal O}(\eps_j^2)$, are
\begin{eqnarray}
P(\nu_e\leftrightarrow\nu_\mu) &\simeq&
\eps_3^2 \left[ 2\cos^2\theta (\sin^2\Delta_{21}
+ \sin^2\Delta_{31}) - \sin^2\Delta_{32} \right. \nonumber\\
&& {}+ \left. 2\sin^2\theta( \sin^2\Delta_{20}
+ \sin^2\Delta_{30}) - \sin^22\theta\sin^2\Delta_{01} \right] \;,
\label{prob1} \\
P(\nu_e\leftrightarrow\nu_\tau) &\simeq& \eps_3^2 \sin^2\Delta_{32} \;,
\label{prob2} \\
P(\nu_e\leftrightarrow\nu_s) &\simeq& \sin^22\theta \sin^2\Delta_{01} \;,
\label{prob3} \\
P(\nu_\mu\leftrightarrow\nu_\tau) &\simeq& \sin^2\Delta_{32} \;,
\label{prob4} \\
P(\nu_\mu\leftrightarrow\nu_s) &\simeq&
\sin^22\theta\eps_3^2 \sin^2\Delta_{01} \;,
\label{prob5}
\end{eqnarray}
where $\Delta_{01} \ll \Delta_{32} \ll \Delta_{20} \simeq \Delta_{30}
\simeq \Delta_{21} \simeq \Delta_{31}$ due to the neutrino mass spectrum.
In our model, only the \nutau--\nus\ oscillation is suppressed beyond
${\cal O}(\eps_j^2)$.

\subsection{Short baseline}

For small $L/E$ only the leading oscillations
$\Delta_{20} \simeq \Delta_{21}
\simeq \Delta_{30} \simeq \Delta_{31}$ contribute, and the only
appreciable oscillation probability is
\begin{equation}
P(\nu_e\leftrightarrow\nu_\mu) \simeq 4\eps_3^2 \sin^2\Delta \;,
\label{lsndosc}
\end{equation}
where $\Delta\equiv m^2 L/4E$. From Eq.\ (\ref{lsndosc}) we can fix
two model parameters
\begin{equation}
\delta m^2_{sbl} = m^2 \;, \quad A^{\mu e}_{sbl} = 4 \eps_3^2 \;.
\label{lsndcon}
\end{equation}
Since the only short--baseline oscillation is
$\nu_e\leftrightarrow\nu_\mu$, these models predict the equality of the
$\nu_e$ disappearance probability, the $\nu_\mu$ disappearance
probability, and the LSND $\nu_\mu\rightarrow\nu_e$ appearance
probability in short--baseline experiments.

\subsection{Long baseline}

For $L/E$ typical to atmospheric or long baseline neutrino experiments,
the oscillations in $\Delta$ assume their average values. The $\Delta_{32}$
oscillation is now evident, and the non-negligible oscillation
probabilities in vacuum are
\begin{equation}
P(\nu_\mu\leftrightarrow\nu_\tau) \simeq \sin^2\Delta_{32} \;,
\label{lb1}
\end{equation}
\begin{equation}
P(\nu_e\leftrightarrow\nu_\mu) \simeq
\eps_3^2 \left(2 - \sin^2\Delta_{32}\right) \;,
\label{lb2}
\end{equation}
\begin{equation}
P(\nu_e\leftrightarrow\nu_{\tau}) \simeq \eps_3^2 \sin^2\Delta_{32} \;.
\label{lb3}
\end{equation}
{}From Eq.\ (\ref{lb1})
\begin{equation}
\delta m^2_{atm} = \delta m^2_{32} \simeq 4 m^2 \eps_4 \;, \quad
A^{\mu\not\mu}_{atm} = 1 \;,
\label{atmpar}
\end{equation}
which determines another parameter of the model.
The model automatically gives maximal oscillations for atmospheric
$\nu_\mu$'s, while oscillations in other channels are suppressed. The
$\nu_\mu$ maximal mixing is natural in the sense that it results from
the large value of the $M_{\nu_\mu \nu_\tau}$ matrix element relative to the
diagonal $M_{\nu_\mu \nu_\mu}$ and $M_{\nu_\tau \nu_\tau}$ elements,
without any need for fine tuning of the difference
$|M_{\nu_\mu \nu_\mu}-M_{\nu_\tau \nu_\tau}|$.

\subsection{Extraterrestrial baseline}

Finally, for very large $L/E \gg
(\delta m^2_{atm}/{\rm eV}^2)^{-1}$~km/GeV, $\sin^2\Delta_{32}$
averages to ${1\over2}$ and the appreciable oscillations in vacuum are
\begin{eqnarray}
P(\nu_e\leftrightarrow\nu_s) \simeq &
\sin^22\theta \sin^2\Delta_{01} \,,
\\
P(\nu_\mu\leftrightarrow\nu_s) \simeq &
\eps_3^2 \sin^22\theta \sin^2\Delta_{01} \;,
\\
P(\nu_e\leftrightarrow\nu_\mu) \simeq &
\eps_3^2 \left[{3\over2} - \sin^22\theta\sin^2\Delta_{01}\right]\,,
\\
P(\nu_e\leftrightarrow\nu_\tau) \simeq &
{1\over2} \eps_3^2 \,,
\\
P(\nu_\mu\leftrightarrow\nu_\tau) \simeq &
{1\over2} \;.
\end{eqnarray}
The solar data can then be explained if the parameters in vacuum satisfy
\begin{eqnarray}
{\rm SAM}: & \delta m^2_{sun} =
\delta m^2_{01} \simeq m^2 \eps_1^2 \;, \quad
A^{e\not e}_{sun} = {4 \eps_2^2 \over 4 \eps_2^2 + \eps_1^2}
\simeq {4 \eps_2^2\over\eps_1^2} \;,
\label{SAMpar}
\\
{\rm LAM}: & \delta m^2_{sun} =
\delta m^2_{01} \simeq
m^2 \eps_1 \sqrt{\eps_1^2 + 4 \eps_2^2} \;, \quad
A^{e\not e}_{sun} = {4 \eps_2^2 \over 4 \eps_2^2 + \eps_1^2} \;,
\label{LAMpar}
\\
{\rm VLW}: & \delta m^2_{sun} =
\delta m^2_{01} \simeq
2 m^2 \eps_2 \eps_3^2 \eps_4 \;, \quad A^{e\not e}_{sun}
= {4 \eps_2^2 \over 4 \eps_2^2 + \eps_1^2} \simeq 1 \;,
\label{VLWpar}
\end{eqnarray}
in the three cases.

\subsection{Determination of the parameters}

In any of these scenarios in Sec.~5.1-5.3, the parameters $m$, $\eps_2$,
$\eps_3$, $\eps_4$, and
$\eps_1$ are obtained from the data in exactly the
same way, i.e., via Eqs.\ (\ref{lsndcon}), (\ref{atmpar}), and
(\ref{SAMpar})-(\ref{VLWpar}). This is a consequence of the
$\nu_s$--$\nu_\tau$ ambiguity. In all cases, the parameters $m$,
$\eps_3$, and $\eps_4$ are related to the observables by
\begin{equation}
m^2 = \delta m^2_{sbl} \,, \quad
\eps_3^2 = {A^{\mu e}_{sbl}\over4} \,, \quad
\eps_4 = {\delta m^2_{atm} \over 4\delta m^2_{sbl}} \; .
\end{equation}
In the solar sector we have
\begin{eqnarray}
{\rm SAM}: & \eps_1^2 =
{\delta m^2_{sun}\over \delta m^2_{sbl}} \,, \quad
\eps_2^2 = {A^{e\not e}_{sun}\delta m^2_{sun}\over 4\delta m^2_{sbl}}\, \,,
\\
{\rm LAM}: & \eps_1^2 =
{\delta m^2_{sun} \sqrt{1 - A^{e\not e}_{sun}} \over
\delta m^2_{sbl}} \,, \quad
\eps_2^2 = {\delta m^2_{sun} A^{e\not e}_{sun} \over
4 \delta m^2_{sbl} \sqrt{1-A^{e\not e}_{sun}}} \, \,,
\\
{\rm VLW}: & \eps_1 \simeq 0 \,, \quad
\eps_2 = {8\delta m^2_{sun}\over A^{\mu e}_{sbl} \delta m^2_{atm}}
\, \,.
\end{eqnarray}

For the specific values $\delta m^2_{sbl}= 2{\rm~eV}^2$ and
$A^{\mu e}_{sbl} = 2.5\times10^{-3}$, $\delta m^2_{atm} =
5\times10^{-3}\rm~eV^2$ and $A^{\mu\not\mu}_{atm} = 1$, and
$\delta m^2_{\rm sol.} = 4\times10^{-6}\rm~eV^2$ and
$A^{e\not e}_{sun} = 1\times10^{-2}$, the
model parameters are given in Table~\ref{params1}.
If we take instead $\delta m^2_{sbl} = 0.3{\rm~eV}^2$
and $A^{\mu e}_{sbl} = 4.0\times10^{-2}$ (which gives the
smallest value of $\delta m^2_{sbl}$ allowed by the data), we get
the model parameters shown in Table~\ref{params2}.
In either of these two examples, the $\delta m^2$ scale for the
atmospheric neutrino oscillation can be adjusted simply by varying
$\eps_4$.  Also in either case, the two heaviest masses provide relic
neutrino targets for a mechanism that may generate the cosmic ray air
showers observed above $\gsim 10^{20}$~eV~\cite{relic}. We note that the
model parameters in Tables~\ref{params1} and \ref{params2} obey the
hierarchies described in Eqs.~(\ref{ordera})-(\ref{orderc}).

\section{Model predictions}

\subsection{Resolving the $\nu_\tau$ vs. $\nu_s$ ambiguity}

If the solar oscillations are $\nu_e\rightarrow\nu_s$ as described in
Sec.~5.1, then our four-neutrino model predicts that the atmospheric
oscillations are $\nu_\mu\rightarrow\nu_\tau$. On the other hand if the
solar oscillations are $\nu_e\rightarrow\nu_\tau$ as in Sec.~5.2, the
atmospheric oscillations are $\nu_\mu\rightarrow\nu_s$.  Several
possibilities have been discussed to resolve the ambiguous assignment of
$\nu_\tau$ and $\nu_s$ as the oscillation partners of the $\nu_e$'s in
the sun and the $\nu_\mu$'s in the atmosphere.  The Solar Neutrino
Observatory (SNO)~\cite{SNO}, which can measure both charge-current (CC)
and neutral-current (NC) interactions, will be able to test whether the
solar $\nu_e$'s oscillate to sterile or active neutrinos: in the sterile
case the CC/NC ratio in SNO will be unity and both CC and NC rates will
be suppressed from the SSM predictions, while in the active case only
the CC rate is suppressed. Of course if the CC measurement is consistent
with the NC, one needs additional evidence to rule out the possibility
that the SSM is in error. For instance, SuperKamiokande and SNO can also
accurately measure the shape of the $^8$B neutrino spectrum, which would
be distorted by oscillations. Also, a measurement of lower energy
neutrinos, such as by the BOREXINO experiment~\cite{borexino}, could
also be used to detect deviations from the SSM spectrum.

Turning to the atmospheric data, the possibilities to resolve
the ambiguity center around the earth--matter effects which are
possible in the \numu\--\nus\ oscillation channel but not in the
\numu\--\nutau\ channel; there is a relative phase difference
between $\nu_\mu$ and $\nu_s$ due to neutral current forward scattering,
but there is no phase difference between $\nu_\mu$ and $\nu_\tau$. The
analytical analysis of matter--effects involving active and sterile
neutrinos can be somewhat complicated~\cite{ls97}, but the
Schr\"odinger--like evolution equations can always be solved
numerically~\cite{bfb}. Other tests have been proposed recently to
resolve the $\nu_\tau$--$\nu_s$ ambiguity. One test is to measure the
asymmetry between downward-going and upward-coming events, for electrons
and muons separately~\cite{flp98}. Various oscillation scenarios give
rise to dramatically differing trajectories of the asymmetries versus
energy for muons and electrons. The preliminary data from SuperK for the
individual muon and electron asymmetries suggests again that the
atmospheric anomaly is primarily due to $\nu_\mu$ oscillating into
either $\nu_\tau$ or $\nu_s$, but not $\nu_e$.  By eventually measuring
an up-down asymmetry for neutral current (NC) events
(e.g. $\nu N\rightarrow\nu N\pi^0$), the ambiguity can be resolved: for
the $\nu_\tau$ case there is no NC asymmetry, whereas for the $\nu_s$
case there is a large NC asymmetry, as shown in Ref.~\cite{lps98}. The
ratio of the rates of NC events relative to the charged current (CC)
events can be also used to the same end~\cite{vs97}, as can multi-ring
events~\cite{hall98}. Searches for muon-less events which come from
$\nu_\tau$, in association with a $\nu_\mu$ disappearance measurement,
can also in principle distinguish between $\nu_\tau$ and
$\nu_s$~\cite{atmnew}.

\subsection{New oscillation signals}

Assuming that the solar oscillations are $\nu_e\rightarrow\nu_s$, we can
determine the new oscillation signals predicted by the model. Given the
order of magnitude of the $\delta m^2_{ij}$ and $U_{\alpha j}$,
observable new phenomenology occurs for $L/E \gg 1$~km/GeV in the
oscillation channels
\begin{eqnarray}
P(\nu_e \leftrightarrow \nu_\mu) &\simeq&
{1\over4} A^{\mu e}_{sbl} (2-\sin^2\Delta_{atm}) \;,
\label{emu}\\
P(\nu_e \leftrightarrow \nu_\tau) &\simeq&
{1\over4} A^{\mu e}_{sbl} \sin^2\Delta_{atm} \;,
\label{etau}
\end{eqnarray}
where $A^{\mu e}_{sbl}\sim {\cal O}(1\%)$ is the oscillation 
amplitude which describes the LSND results and $\Delta_{atm}
= 1.27 \delta m^2_{atm} L/E \sim
(\delta m^2_{atm}/5\times10^{-3}{\rm~eV}^2)
(L/157{\rm km})({\rm GeV}/E)$ is the oscillation argument
which describes the atmospheric neutrino data.
We emphasize the new predictions in the $\nu_e \leftrightarrow \nu_\mu$ and
$\nu_e \leftrightarrow \nu_\tau$ channels:
long baseline oscillations with common oscillation length
determined by the {\it atmospheric} $\Delta_{atm}$ and
common amplitude given by ${1\over4}$ times the {\it LSND} amplitude
$A^{\mu e}_{sbl}$.  These oscillations are in addition to the
$\nu_\mu \leftrightarrow \nu_e$ oscillations due to $\Delta_{sbl}$ in
Eq.\ (\ref{lsndosc}), which average to the value
of ${1\over2} A^{\mu e}_{sbl}$ in a long baseline experiment.
The amplitudes and lengths of these new oscillations complement the
set in Table~\ref{amplength}, which are inevitable, given the present data,
and are therefore required in any model.


How can the oscillation probabilities in Eqs.\ (\ref{emu}) and (\ref{etau})
be tested? A list of experiments currently underway or
being planned to test neutrino oscillation hypotheses is given in
Table~\ref{oscexp}~\cite{experiments}. In each case the oscillation channel
and the parameters which are expected to be tested are shown.

The MINOS experiment \cite{MINOS} can detect
$\nu_\mu\rightarrow\nu_e$ or $\nu_\mu\rightarrow\nu_\tau$ oscillations
and is sensitive down to $\delta m^2 \simeq 10^{-3}$~eV$^2$ and a mixing
amplitude of $10^{-2}$, which partially overlaps the region of interest;
see Fig.~8. If the MINOS experiment can increase its sensitivity, it
will provide an even better test of this new phenomenology.

Long--baseline experiments with an intense $\nu_e$ or $\bar\nu_e$
neutrino beam and which can detect $\tau$'s can see the
$\nu_e\rightarrow\nu_\tau$ oscillations in Eq.\ (\ref{etau}) and provide
a definitive test of the new phenomenology predicted by the model.
High intensity muon sources~\cite{Geer} can
provide simultaneous high intensity $\nu_\mu$ and $\bar\nu_e$ (or
$\bar\nu_\mu$ and $\nu_e$ for antimuons) beams with well--determined
fluxes, which could then be aimed at a neutrino detector at a distant
site.  It is expected that $\tau$'s will be detected through their
$\mu$ decay mode and that a charge determination can be made, so that
one can tell if the $\tau$ originated from $\nu_\mu \rightarrow \nu_\tau$
or $\bar\nu_e \rightarrow \bar\nu_\tau$ oscillations. Current
proposals~\cite{Geer} consider SOUDAN ($L=$~732~km) or GRAN SASSO
($L=$~9900~km) as the far site from an intense muon source at Fermilab
(MC). These experiments could also observe $\nu_e \rightarrow \nu_\mu$
oscillations via detection of ``wrong--sign'' muons, i.e., those with
sign opposite to that expected from the $\nu_\mu$ or $\bar\nu_\mu$
source. The neutrino energies are in the 10--50~GeV range. Assuming that
low backgrounds can be achieved, the sensitivity to $\delta m^2$ is
roughly proportional to the inverse square root of the detector size
(given the same neutrino energy spectrum at the source); the $\delta
m^2$ sensitivity does not depend on detector distance $L$ because
although the flux in the detector falls off with $L^2$, the oscillation
argument grows with $L^2$ for small $\delta m^2 L/E$. For 20~GeV muons
at Fermilab and a 10~kT detector at either SOUDAN or GRAN SASSO, the
single--event $\delta m^2$ sensitivity for $\nu_e \rightarrow \nu_\tau$
oscillations is about $8\times10^{-5}$~eV$^2$ for maximal
mixing~\cite{Geer}. For large $\delta m^2$, the oscillation amplitude
single--event sensitivity is roughly inversely proportional to the
neutrino flux at the detector divided by the detector size; about
$6\times10^{-5}$ for SOUDAN and $10^{-2}$ for GRAN SASSO~\cite{Geer}.
In general, the closer detector has comparable $\delta m^2$ sensitivity
but better $A$ sensitivity.

The model predicts $\nu_e\rightarrow\nu_\tau$ oscillations with
amplitude $\frac{1}{4} A_{sbl}$ (which ranges from 0.0006 to 0.01) and
mass--squared difference of $\delta m^2_{atm}$ (which ranges from
$3\times10^{-4}$ to $7\times10^{-3}$ eV$^2$). The region of possible
$\nu_e\rightarrow\nu_\tau$ oscillations in our model and the regions
which can be tested at the SOUDAN and GRAN SASSO sites with a neutrino
beam from a high-intensity muon source at Fermilab are shown
schematically in Fig.~8, along with the favored parameters for the LSND,
atmospheric neutrino, and solar neutrino oscillations. Such experiments
would be sensitive to some of the $\nu_e \rightarrow
\nu_\tau$ region, though they may not cover the low--mass,
small--amplitude part. These searches would also be able to test the
$\nu_e \rightarrow \nu_\mu$ oscillations in Eq.\ (\ref{emu}) and the
atmospheric $\nu_\mu \rightarrow \nu_\tau$ oscillations. Additionally,
long baseline experiments to the AMANDA detector~\cite{AMANDA} from
Fermilab ($L\simeq$ 11700 km) or KEK ($L\simeq$ 11300 km) may be useful in
probing oscillations with small $\delta m^2$.

If the solar oscillations are $\nu_e\rightarrow\nu_\tau$, then the
oscillations of atmospheric neutrinos are $\nu_\mu\rightarrow\nu_s$ and
the new oscillations in Eq.\ (\ref{etau}) are instead
$\nu_e\rightarrow\nu_s$. Neither of these signals would be detectable
in long--baseline experiments since the signal is $\nu_e$ or $\nu_\mu$
disappearance at the few percent level or less. The only measurable
signal of the model in this case is the $\nu_e\rightarrow\nu_\mu$
oscillations in Eq.\ (\ref{emu}).

If the solar neutrinos oscillate into both $\nu_s$ and $\nu_\tau$ as
given by the mixing in Eq.\ (\ref{Uprime}), any vacuum oscillation in
Secs.~6.1-6.4 which has $\nu_s$ as the final state is replaced by
oscillations to $\nu_s$ with relative probability $\cos^2\alpha$ and to
$\nu_\tau$ with relative probability $\sin^2\alpha$. Conversely, any
oscillation which has $\nu_\tau$ as the final state is replaced by
oscillations to $\nu_s$ with relative probability $\sin^2\alpha$ and to
$\nu_\tau$ with relative probability $\cos^2\alpha$. In particular, a
solar $\nu_e$ oscillates into a mixture of $\nu_s$ and $\nu_\tau$ with
relative probability $\cos^2\alpha$ and $\sin^2\alpha$, respectively, in
a vacuum, and an atmospheric $\nu_\mu$ oscillates into a mixture of
$\nu_s$ and $\nu_\tau$ with relative probability $\sin^2\alpha$ and
$\cos^2\alpha$, respectively, in a vacuum. The new oscillation signal
in Eq.\ (\ref{etau}) for long baseline experiments is replaced by
\begin{eqnarray}
P(\nu_e\leftrightarrow\nu_\tau) \simeq &
{1\over4} A^{\mu e}_{sbl} \cos^2\alpha \sin^2\Delta_{atm} \;, \\
P(\nu_e\leftrightarrow\nu_s) \simeq &
{1\over4} A^{\mu e}_{sbl} \sin^2\alpha \sin^2\Delta_{atm} \;.
\label{lbprime}
\end{eqnarray}
Also, there are new oscillations between $\nu_s$ and $\nu_\tau$, with
the general probability for any baseline given by
\begin{eqnarray}
P(\nu_s\rightarrow\nu_\tau) = & \sin^2\alpha\cos^2\alpha
\left[ 2 \cos^2\theta (\sin^2\Delta_{20}+\sin^2\Delta_{30}) \right.
\nonumber \\
& \left. + 2 \sin^2\theta (\sin^2\Delta_{21}+\sin^2\Delta_{31})
-\sin^2\Delta_{32}-\sin^22\theta\sin^2\Delta_{01} \right] \;.
\end{eqnarray}

\subsection{Neutrinoless double--$\beta$ decay}

Since the $M_{\nu_e\nu_e}$ mass matrix element is zero in Eq.\ (\ref{m}),
there is no neutrinoless double-$\beta$ decay at tree level in our
model.  The present limit on $M_{\nu_e\nu_e}$ from this process is
$\sim$0.5 eV~\cite{KK}.
New experiments are under development which may measure $M_{\nu_e\nu_e}$
down as low as 0.01 eV\cite{KK}.
If a nonzero $M_{\nu_e\nu_e}$ is found at these levels, it would be
incompatible with the solar solutions in our models.

\subsection{Tritium decay}

If $\nu_e$ is primarily associated with the lighter pair in the 2+2
model, and $m_1 < m_0 \ll m_2, m_3$, then there will be no measurable
effect in the endpoint of the tritium decay spectrum. Since only
mass-squared differences are important for oscillations, the inverted
2+2 model, where $m_2, m_3 \ll m_1 < m_0$, is equally viable, although
it is not derivable from the mass matrix in Sec.~5. Then the
$\nu_e$ will have an effective mass of $0.55-1.4$~eV, which is just
below the current limit~\cite{tritium}.

\subsection{Hot dark matter}

For $m\simeq$~1.4~eV, approximately the largest value allowed by the
LSND data, $\sum m_\nu \simeq 3$~eV, which according to recent work on
early universe formation of the largest structures provides an ideal hot
dark matter component~\cite{hdm}. For $m\simeq$~0.55~eV, the
contribution to hot dark matter is much smaller.  The contribution of
the neutrinos to the mass density of the universe is given by
$\Omega_\nu = \sum m_\nu/(h^2 93$~eV), where $h$ is the Hubble expansion
parameter in units of 100 km/s/Mpc~\cite{expansion}; with $h=0.65$ our
model implies $\Omega_\nu \simeq 0.05$. An interesting test of neutrino
masses is the power spectrum of early galaxy sizes, to be provided by
the Sloan Digital Sky Survey (SDSS)~\cite{SDSS}. For two
nearly degenerate massive neutrino species, sensitivity down to about
0.2 to 0.9~eV (depending on $\Omega$ and $h$) is expected, providing
coverage of all or part of the LSND allowed range ($m=$0.55 to 1.4~eV in
our model). Also, the future MAP~\cite{MAP} and
PLANCK~\cite{PLANCK} satellite missions, which will measure the cosmic
microwave background radiation, should be sensitive to neutrino
densities to high precision \cite{TonT}, and in particular to the
$\nu_\mu\rightarrow\nu_s$ atmospheric or the $\nu_e\rightarrow\nu_s$
large-angle MSW neutrino mixing solutions~\cite{TonTsterile}.

\subsection{Resonant enhancement in matter}

The curves in Fig.~8 (in the scenario where the solar oscillations are
$\nu_e\rightarrow\nu_s$) assume vacuum oscillations in the Earth. In
general, large corrections to oscillations involving $\nu_e$ and $\nu_s$
are possible due to matter.  The $\nu_e$ diagonal element in the
effective mass-squared matrix receives an additional term $\sqrt2 G_F
N_e E$ from its forward elastic CC interaction, and the $\nu_s$ diagonal
element receives the contribution $G_F N_n E/\sqrt2$ (relative to the
active neutrinos) because it does not have NC interactions.  Here, $N_e$
and $N_n$ denote the electron and neutron number density, respectively.
In Table~\ref{Eres} we display the resonant energies in the earth for
the various vacuum $\delta m^2$ values suggested by the available LSND,
atmospheric, and solar data.  For neutrinos with energies significantly
above $E_{res}$, oscillations are suppressed; for neutrinos with
energies significantly below $E_{res}$, the matter effect is negligible;
at the resonant energy, $A_{{\rm mat}}=1$ and the oscillation length is
increased by $1/\sqrt{A_{{\rm vac}}}$.

Some of the resonant energy values in Table~\ref{Eres} are of particular
interest.  Upcoming neutrinos from the atmosphere or astrophysical
sources, with mass at the lower end of the LSND range, can have their
oscillations resonantly enhanced by the earth's mantle and/or
core. Atmospheric neutrinos below a few GeV and the SAM and LAM $pp$
neutrinos from the sun also appear to be near resonance in the earth's
matter. Day--night modulation of the solar flux due to
earth--matter effects is expected to discriminate between
\nutau\ and \nus\ solar fluxes~\cite{petcov98,akhmedov98}, while a
precise measurement of zenith angle dependence may discriminate
between \nutau\ and \nus\ atmospheric fluxes~\cite{lms98}. Oscillation
wavelengths commensurate with the size of the earth's mantle and/or core
are especially sensitive.

In our model of Sec.~5.1 with $\nu_\mu\rightarrow\nu_\tau$ atmospheric
neutrino oscillations, however, these corrections do not significantly
affect the large $m_2^2$ and $m_3^2$ mass eigenvalues as long as $E
\ll$~1~TeV, and hence only modify the $\delta m^2_{01}$ oscillation
argument. We have verified this by explicit diagonalization of the mass
matrix when matter effects are included. Hence we conclude that the
matter corrections to the mass matrix in Eq.~(\ref{m}) probably have no
observable consequences in all terrestrial experiments. For large $L/E$,
such as when $E \lsim 10$~MeV for solar neutrinos, the only significant
effect of matter is the usual MSW enhancement of $\nu_e \rightarrow
\nu_s$ that leads to the solar neutrino suppression of the \nue\ flux;
in all other oscillation channels the matter--enhanced amplitudes are at
the $\eps_{\alpha\beta}^2$ level or smaller. In the models discussed
in Secs.~5.2 and 5.3, where there is a $\nu_\tau-\nu_s$ component to the
atmospheric oscillation, matter effects as discussed here may be
important in terrestrial experiments \cite{ls97}.

\section{Discussion and summary}

\subsection{Distinguishing the three solar solutions}

The VLW solar solution may be discriminated from the two MSW solutions
by a careful measurement of the solar neutrino spectrum by SuperK and
BOREXINO~\cite{spectrum}, or by determining the amount of seasonal
variation of the $^7$Be and $pep$ neutrinos~\cite{bpw90}, which
can be measured by the BOREXINO experiment. The $^8$B neutrino spectrum
as measured in SuperK and SNO will also be useful in discriminating
between the SAM and LAM solutions~\cite{samlam}. 
The HERON and HELLAZ~\cite{hellaz} experiments would be able to
measure the $pp$ neutrino energy spectrum, which would also be useful in
differentiating the three scenarios.

\subsection{Possible $\nu_{\tau}$--$\nu_s$ mixing}

The general case with \nus\--\nutau\ mixing is described in Sec.~5.3.
The unmixed cases Eqs.~(\ref{m}) and (\ref{m2}) are obtained
in the limits $\alpha\rightarrow 0$ and
$\alpha\rightarrow \frac{\pi}{2}$, respectively; distinguishing between
the unmixed scenarios is discussed in Sec~7.1. How might non-trivial
mixing of \nutau\ and \nus\ be observed, and how might the mixing angle
be deduced? A mixed model would generally have a signature intermediate
between the two unmixed signatures~\cite{ls97}; e.g., experiments
measuring neutral current events for solar and atmospheric neutrinos
would find a result between those expected for $\nu_\tau$ and $\nu_s$.

\subsection{Summary}

An analysis of all the available data (short baseline LSND, reactor
and accelerator, long baseline atmospheric, and extraterrestrial length
solar) in terms of neutrino oscillations leads to the conclusion that
three independent oscillation lengths are contributing.
This then further requires mixing of at least four light--mass neutrinos.
For a four light--mass neutrino universe,
we draw the following model--independent conclusions:\\
(i) the 1+3 (or 3+1) mass spectrum with a separated mass is disfavored
when all the data (LSND, reactor, accelerator, solar, and atmospheric)
are considered, leaving a spectrum with two nearly--degenerate pairs
as preferred;\\
(ii) the neutrino mixing matrix elements satisfy $|U_{e2}/U_{e3}|
\simeq |U_{\mu 2}/U_{\mu 3}|$ if $\delta m^2_{sbl} \simeq 0.3$~eV$^2$,
i.e., the parameters lie near the Bugey $\nu_e$ disappearance limit;\\
(iii) the relation $|U_{\mu 2}| \simeq |U_{\mu 3}|$ is inferred from
the near--maximal mixing of atmospheric $\nu_\mu$'s measured
by SuperK, which together with (ii) implies $|U_{e2}| \simeq |U_{e3}|$.

Based upon the apparent need for more than three light neutrinos, we
have presented four-neutrino models with three active neutrinos and one
sterile neutrino.  The models naturally have maximal
$\nu_\mu\rightarrow\nu_\tau$ (or $\nu_\mu\rightarrow\nu_s$) oscillations
of atmospheric neutrinos and can also explain the solar neutrino and
LSND results. The solar solutions can be $\nu_e\rightarrow\nu_s$ or
$\nu_e\rightarrow\nu_\tau$, and can be small-angle matter-enhanced,
large-angle matter-enhanced, or vacuum long--wavelength oscillations;
the increased statistics on the electron energy distribution and
day/night differences of the SuperK data \cite{newSK} may further
clarify the allowed regions for the solar solutions. 
The models predict $\nu_e\leftrightarrow \nu_\tau$
(or $\nu_e\rightarrow\nu_s$) and $\nu_e \leftrightarrow \nu_\mu$
oscillations in long--baseline experiments with $L/E \gg 1$~km/GeV with
amplitudes that are determined by the LSND oscillation amplitude and
$\delta m^2$ scale determined by the oscillation scale of atmospheric
neutrinos. For the $\nu_e\rightarrow\nu_\tau$ case, these oscillations
might be seen by experiments based on neutrino beams from an intense
muon source at Fermilab with a detector at the SOUDAN or GRAN SASSO
sites. The $\nu_\mu\rightarrow\nu_e$ oscillations might be seen by the
MINOS experiment or at KEK with detectors at Kamiokande and
SuperKamiokande. The models also predict the equality of the $\nu_e$
disappearance probability, the $\nu_\mu$ disappearance probability, and
the LSND $\nu_\mu\rightarrow\nu_e$ appearance probability in
short--baseline experiments.

\section{Acknowledgements}

We thank D. Caldwell, R. Foot, P. Krastev, J.G. Learned,
S. Petcov, G. Raffelt, J. Stone, and R. Volkas for comments and discussions.
This work was supported in part by the U.S. Department of Energy, Division of
High Energy Physics, under Grants No.~DE-FG02-94ER40817,
No.~DE-FG05-85ER40226, and No.~DE-FG02-95ER40896, and in part by the
University of Wisconsin Research Committee with funds granted by the
Wisconsin Alumni Research Foundation,
the Vanderbilt University Research Council, the University of Hawaii,
and the Max Planck Institute for Physics, Munich.

\vfill\eject

\newpage


\begin{table}
\caption[]{Ranges of mass-squared differences and amplitudes that provide
oscillation solutions to the solar neutrino data within 95\%~C.L. in
the small--angle MSW, large--angle MSW and vacuum long--wavelenth
scenarios when $\nu_e\rightarrow\nu_s$ or
$\nu_e\rightarrow\nu_\tau$~\cite{hata}. The new SuperK data and new
solar flux calculations give slightly different oscillation parameters
than those quoted here; in particular, the $\delta m^2$ values for the
VLW solution are higher, of order $4\times10^{-10}$~eV$^2$
\cite{totsuka}.}
\label{sunsol}
\vspace{0.5 cm}
\vbox{\footnotesize
\begin{tabular}{|c|c|c|c|}
\hline
$\nu_e\rightarrow\nu_s$ & SAM & LAM & VLW \\ \hline
$\delta m^2$ (eV$^2$) & $3.5\times10^{-6}$--$7.5\times10^{-6}$
& $\simeq9\times10^{-6}$ & $\simeq5\times10^{-11}$ \\
$A^{es}_{sun}$ & $2.5\times10^{-3}$--$1.6\times10^{-2}$
& $\simeq0.7$ & $\simeq1$ \\
\hline
$\nu_e\rightarrow\nu_\tau$ & SAM & LAM & VLW \\
\hline
$\delta m^2$ (eV$^2$) & $4\times10^{-6}$--$9\times10^{-6}$
& $8\times10^{-6}$--$3\times10^{-5}$
& $5\times10^{-11}$--$8\times10^{11}$ \\
$A^{e\tau}_{sun}$ & $3.5\times10^{-3}$--$1.3\times10^{-2}$
& $0.4$--$0.8$ & $0.6$--$1.0$ \\
\hline
\end{tabular} }
\end{table}


\begin{table}
\caption[]{Some typical $\delta m^2$ and vacuum oscillation amplitudes
suggested by experiment, and the corresponding vacuum oscillation lengths
$\lambda_v=2.5\,(E_\nu/{\rm GeV})(\delta m^2/{\rm eV}^2)^{-1}$ km
for representative neutrino energies.
Here AU is the earth-sun distance and $R_\oplus$ is the earth's radius.}
%
\label{amplength}
\vspace{0.5 cm}
\vbox{\footnotesize
\begin{tabular}{|c c|c|c|c|c|c|c|c|c|}
\hline
 & $\delta m^2/{\rm eV}^2$ & $A$ &
     \multicolumn{7}{c}
           {$\lambda_v=4\pi E_\nu/\delta m^2$, with $E_\nu$ as shown:}\\
&&& 5  MeV & 100 MeV & 2 GeV & 10 GeV & 30 GeV & 100 GeV
          & 1 TeV\\ \hline
LSND & 2 & 0.0025 & 6 m & 125 m & 2.5 km & 12 km & 37 km
          & 125 km & 1250km \\ \hline
LSND & 0.3 & 0.04 & 42 m & 840 m & 17 km & 83 km & 250 km & 830 km
     & 1.2 $R_\oplus$\\ \hline
ATM & $5\times 10^{-3}$ & 0.8--1.0 & 2.5 km & 50 km & $10^3$ km
    & 5000km &
    2.3 $R_\oplus$ & 7.8 $R_\oplus$ & 78 $R_\oplus$\\ \hline
SAM & $6\times 10^{-6}$ & .0025--.016 & 2100 km
        & 6.5 $R_{\oplus}$ & 130 $R_\oplus$ &&&&\\ \hline
LAM & $10^{-5}$ & 0.4--0.8 & 1260 km & 3.9 $R_{\oplus}$ & 78 $R_\oplus$
        & & & &\\ \hline
VLW & $5\times 10^{-11}$ & 0.6--1.0  & 1.7 AU & 33 AU & 670 AU
        & & & &\\ \hline
\end{tabular}
}
\end{table}

\newpage


\begin{table}
\caption[]{Typical model parameters for the input data
$\delta m^2_{sbl}=2$~eV$^2$, $A^{\mu e}_{sbl}=2.5\times10^{-3}$,
$\delta m^2_{atm}=5\times10^{-3}$~eV$^2$, and $A^{\mu\not\mu}_{atm}=1$.}
\label{params1}
\vspace{0.5 cm}
\vbox{\footnotesize
\begin{center}
\begin{tabular}{|c|c|c|c|}
\hline
inputs & SAM & LAM & VLW \\
\hline
$\delta m^2_{sun}$ (eV$^2$)
& $4\times10^{-6}$ & $9\times10^{-6}$ & $5\times10^{-11}$ \\
$A^{e\not e}_{sun}$ & $1\times10^{-2}$ & $0.72$ & $1.0$ \\
\hline
outputs & SAM & LAM & VLW \\
\hline
$m$ (eV) & $1.4$ & $1.4$ & $1.4$ \\
$\eps_3$
& $2.5\times10^{-2}$ & $2.5\times10^{-2}$ & $2.5\times10^{-2}$ \\
$\eps_4$
& $6.3\times10^{-4}$ & $6.3\times10^{-4}$ & $6.3\times10^{-4}$ \\
$\eps_2$
& $7.1\times10^{-5}$ & $1.2\times10^{-3}$ & $3.2\times10^{-5}$ \\
$\eps_1$
& $1.4\times10^{-3}$ & $1.5\times10^{-3}$ & $\simeq 0$ \\
\hline
\end{tabular}
\end{center}
}
\end{table}


\begin{table}
\caption[]{Typical model parameters for the input data
$\delta m^2_{sbl}=0.3$~eV$^2$, $A^{\mu e}_{sbl}=4.0\times10^{-2}$,
$\delta m^2_{atm}=5\times10^{-3}$~eV$^2$, and $A^{\mu\not\mu}_{atm}=1$.}
\label{params2}
\vspace{0.5 cm}
\vbox{\footnotesize
\begin{center}
\begin{tabular}{|c|c|c|c|}
\hline
inputs & SAM & LAM & VLW \\
\hline
$\delta m^2_{sun}$ (eV$^2$)
& $4\times10^{-6}$ & $9\times10^{-6}$ & $5\times10^{-11}$ \\
$A^{e\not e}_{sun}$ & $1\times10^{-2}$ & $0.72$ & $1.0$ \\
\hline
outputs & SAM & LAM & VLW \\
\hline
$m$ (eV) & $0.55$ & $0.55$ & $0.55$ \\
$\eps_3$
& $1.0\times10^{-1}$ & $1.0\times10^{-1}$ & $1.0\times10^{-1}$ \\
$\eps_4$
& $4.2\times10^{-3}$ & $4.2\times10^{-3}$ & $4.2\times10^{-3}$ \\
$\eps_2$
& $1.9\times10^{-4}$ & $3.2\times10^{-3}$ & $2.0\times10^{-6}$ \\
$\eps_1$
& $3.7\times10^{-3}$ & $4.0\times10^{-3}$ & $\simeq 0$ \\
\hline
\end{tabular}
\end{center}
}
\end{table}

\newpage


\begin{table}
\caption[]{Current and planned short and long baseline neutrino
oscillation experiments. Stars denote accessible oscillation channels.
The $\delta m^2$ and $\sin^22\theta$ sensitivies are given.}
\label{oscexp}
\def\ds{\displaystyle}
\vbox{\footnotesize
\tabcolsep=.5em
\begin{tabular}{lcccccccccc}
&&&&&&&&& \multicolumn{2}{c}{Test Model?}\\
Experiment&
$\ds\nu_\mu\atop{\ds\downarrow\atop\ds\nu_e}$&
$\ds\nu_\mu\atop{\ds\downarrow\atop\ds\nu_\tau}$&
$\ds\nu_e\atop{\ds\downarrow\atop\ds\nu_\tau}$&
$\ds\nu_e\atop{\ds\downarrow\atop\ds\nu_e}$&
$\ds\delta m^2\atop\bigskip\rm (eV^2)$& $\sin^22\theta$&
$\ds\rm Test\atop\ds\rm LSND?$&$\ds\rm Test\atop\ds\rm Atmos?$&
$\ds\nu_\mu\atop{\ds\downarrow\atop\ds\nu_e}$&
$\ds\nu_e\atop{\ds\downarrow\atop\ds\nu_\tau}$
\\[3mm]
\hline
BOONE&  $\star$& & & & $10^{-2}$& $6\times10^{-4}$&
$\star$&  & &   \\
BOREXINO& & & &  $\star$& $10^{-6}$& 0.4& & & &  \\
CHORUS& & $\star$& & & 0.3& $2\times10^{-4}$& & & & \\
COSMOS& & $\star$& & & 0.1& $10^{-5}$& &  & &   \\
ICARUS, NOE,& $\star$& $\star$& & & $3\times10^{-3}$& $4\times10^{-2}$&  & p&
\\
AQUA-RICH, OPERA\\
KARMEN& $\star$& & & & $4\times10^{-2}$& $10^{-3}$& $\star$& &  &  \\
KamLAND&  $\star$& & & & $2\times10^{-3}$& 0.2&  & &   & \\
K2K& $\star$& & & & $2\times10^{-3}$& $5\times10^{-2}$ & & p \\
MC/Gran Sasso& $\star$& $\star$& $\star$& & $8\times10^{-5}$& $10^{-2}$&
p& $\star$& p& p \\
MC/Soudan& $\star$& $\star$& $\star$& & $8\times10^{-5}$&
$6\times10^{-5}$& $\star$& $\star$& $\star$& p \\
MINOS& $\star$& $\star$& & & $10^{-3}$& $10^{-2}$& p& p& p&   \\
NOMAD& & $\star$& & & 0.5& $3\times10^{-4}$& &  & &   \\
ORLANDO, ESS& $\star$& & & & $3\times10^{-3}$& $10^{-4}$& $\star$& &  &  \\
Palo Verde & & & & $\star$& $10^{-3}$& 0.2&  & & & \\
TOSCA& & $\star$& & & 0.1& $10^{-5}$&\\
\hline
\end{tabular}
\bigskip

 p = partially
}
\end{table}


\begin{table}
\caption[]{Resonant energies
$E_{res}=6.6\,\sqrt{1-A}\,(\delta m^2/{\rm eV}^2)(N_e/N_A \, 
{\rm cm}^{-3})^{-1}$ TeV in the earth's core and mantle
for $\nu_e$--$\nu_{\mu/\tau}$ oscillations,
for some typical vacuum values for the $\delta m^2$'s
and amplitudes suggested by the data. Here $A$ is the oscillation
amplitude and $N_A$ is Avagadro's number.
We have taken the core electron density to be 4.5 to 6.0 $N_A/{\rm cm}^3$,
and the mantle density to be 1.6 to 2.6 $N_A/{\rm cm}^3$.
Resonant energies for $\nu_e$--$\nu_s$ oscillations
are $2N_e/(2N_e-N_n)$ times larger than for
$\nu_e$--$\nu_{\mu/\tau}$ oscillations,
resonant energies for $\nu_{\mu/\tau}$--$\nu_s$ oscillations are
$2N_e/N_n$ times larger, but resonance occurs for ${\bar \nu}$
rather than $\nu$ since the phase difference due to matter has the
opposite sign.}
\label{Eres}
\vspace{0.5 cm}
\centering\leavevmode
\begin{tabular}{|c|c c|c|c|}
\hline
 $A$ & & $\delta m^2/{\rm eV}^2$ &
     \multicolumn{2}{c|}{$E_{res}=\delta m^2\sqrt{1-A}/(2\sqrt{2} G_F N_e)$}\\
 & & & core & mantle\\ \hline
 $\ll 1$ & LSND & 2               & 2.2--2.9 TeV & 5.1--8.2 TeV\\ \hline
 $\ll 1$ & LSND & 0.3             & 330-440 GeV    & 0.8-1.2 TeV\\ \hline
 0.8 & ATM & $5\times 10^{-3}$    & 2.4--3.3 GeV & 5.6--9.2  GeV\\ \hline
 $\ll 1$ & SAM & $6\times 10^{-6}$  & 6.5--8.8 MeV & 15--25 MeV\\ \hline
 0.6 & LAM & $10^{-5}$            & 6.9--9.2 MeV & 16--26 MeV\\ \hline
 0.8 & VLW & $5\times 10^{-10}$   & 25--33 eV  & 55--90 eV \\  \hline
\end{tabular}
\end{table}

\clearpage


\begin{figure}
\centering\leavevmode
\epsfxsize=5in\epsffile{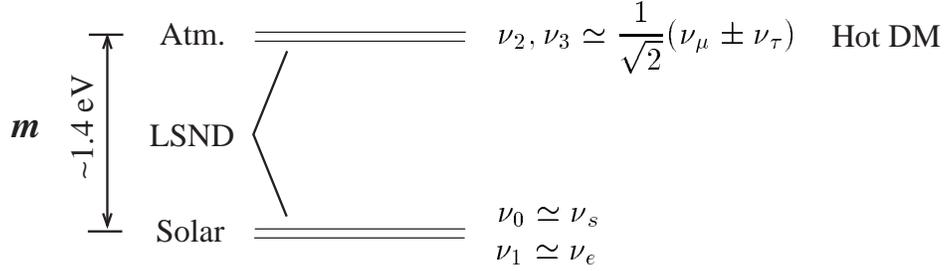}

\caption[]{Neutrino mass spectrum, showing a possible flavor assignment
for each mass eigenstate, and
showing which mass splittings are responsible for the LSND,
atmospheric, and solar oscillations.}
\end{figure}


\begin{figure}
\centering\leavevmode
\epsfxsize=5.5in\epsffile{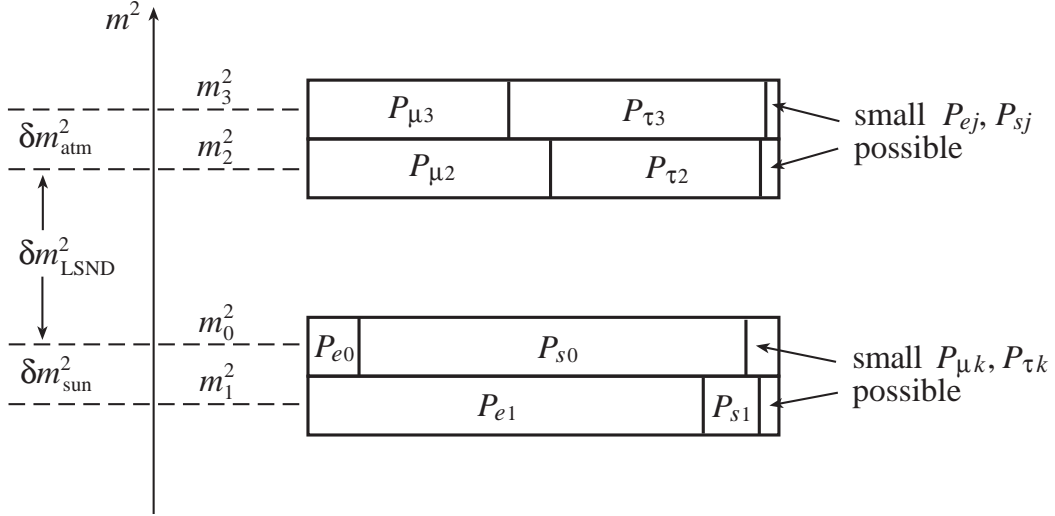}

\caption[]{Typical probability rectangles for a 2+2 model where
$\nu_2$ and $\nu_3$ generate the atmospheric oscillations and
$\nu_1$ and $\nu_0$ generate the solar oscillations. For the non-dominant
probabilities $j=2$ or $3$ and $k=1$ or $0$.}
\end{figure}


\begin{figure}
\centering\leavevmode
\epsfxsize=4in\epsffile{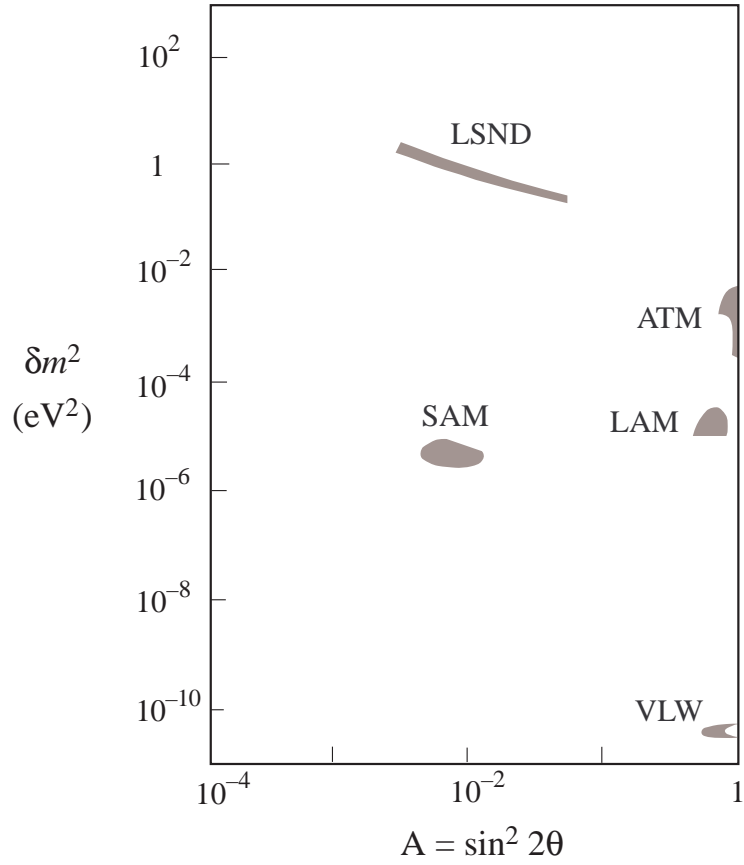}

\caption[]{The three allowed two-neutrino solar solutions for
$\nu_e\rightarrow\nu_\tau$ oscillations~\cite{MSWconf}. The
corresponding region for $\nu_e\rightarrow\nu_s$ oscillations are
similar to the $\nu_e\rightarrow\nu_\tau$ case.}
\end{figure}


\begin{figure}
\centering\leavevmode
\epsfxsize=5.5in\epsffile{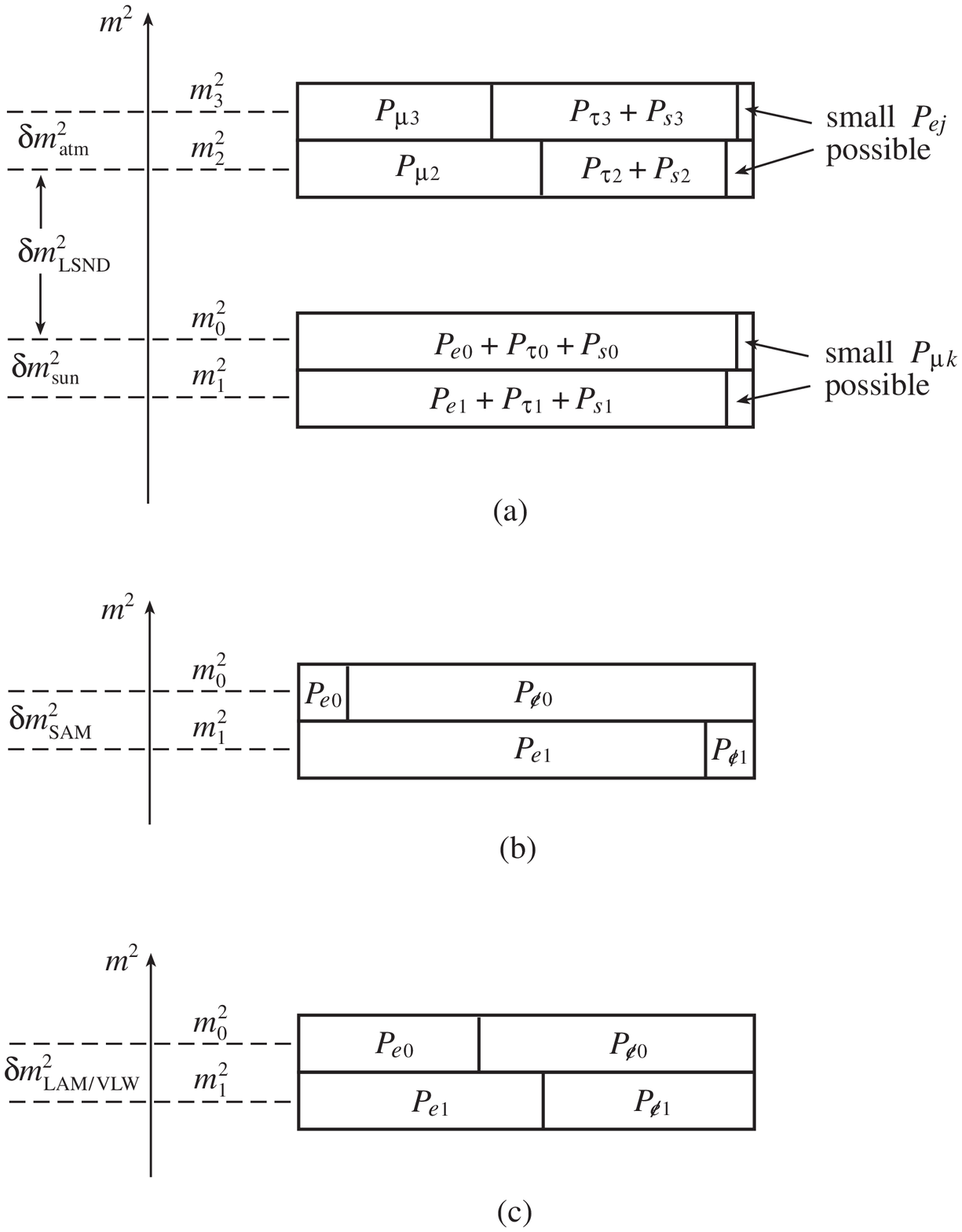}

\caption[]{Typical probability rectangles for some 2+2 models. (a)
Rectangles deduced from atmospheric and CHOOZ neutrino data, with
$\nu_\mu$ lying predominantly in the $\nu_2$ and $\nu_3$
states with large mixing with either $\nu_\tau$ and/or $\nu_s$. The
$P_{\mu 0}$ and $P_{\mu 1}$ probabilites are small, but the $P_{\not\mu
0}$ and $P_{\not\mu 1}$ probabilities are not yet determined.
The partitioning of the $P_{\not\mu 0}$ and $P_{\not\mu 1}$ probabilites
once the solar solution is specified are also shown for the (b) small
angle MSW and (c) large angle MSW or vacuum long--wavelength solutions.}
\end{figure}


\begin{figure}
\centering\leavevmode
\epsfxsize=5.5in\epsffile{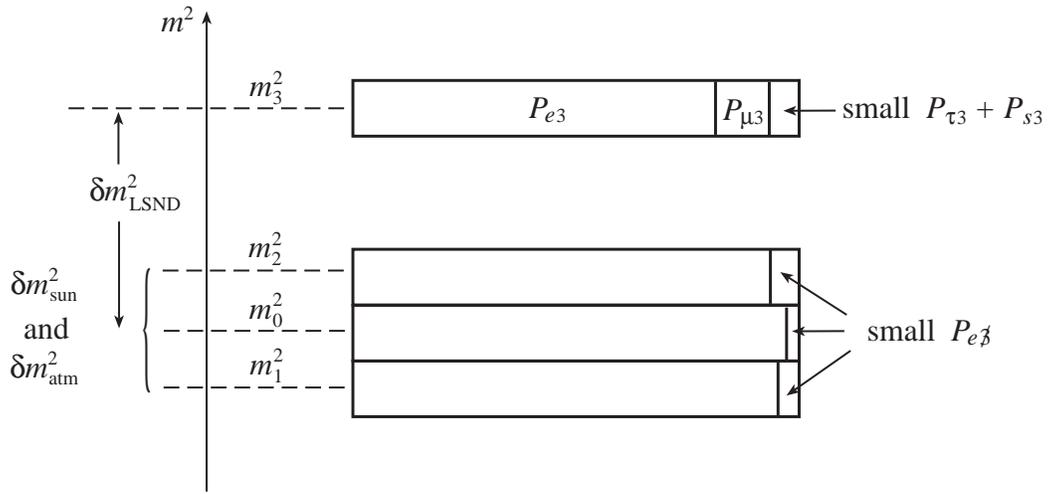}

\caption[]{Typical probability rectangles for the 1+3 model when
$P_{e3}$ is large and $P_{e\not 3}$ is small.}
\end{figure}


\begin{figure}
\centering\leavevmode
\epsfysize=7.5in\epsffile{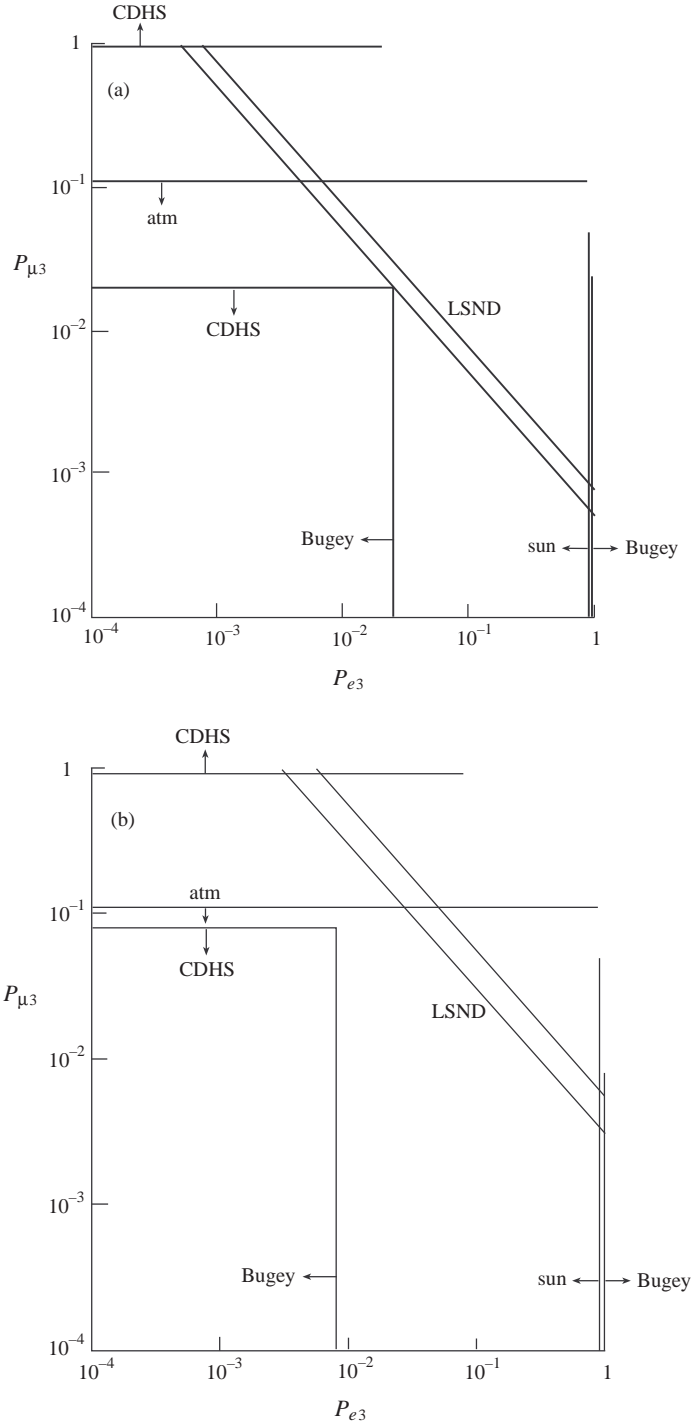}

\caption[]{Constraints on $P_{\mu 3}$ and $P_{e 3}$ for the 1+3 model
when (a) $\delta m^2_{sbl}=1.7$~eV$^2$ and
(b) $\delta m^2_{sbl}=0.5$~eV$^2$. The
LSND, Bugey, CDHS, atmospheric, and solar constraints are obtained by
comparison of the appropriate data with Eqs.~(\ref{emu1}), (\ref{enote}),
(\ref{munotmu}), (\ref{atmbound}), and (\ref{sunbound}), respectively.
Not displayed is the unitarity constraint $P_{e3} + P_{\mu 3} \leq 1$.}
\end{figure}


\begin{figure}
\centering\leavevmode
\epsfxsize=4.5in\epsffile{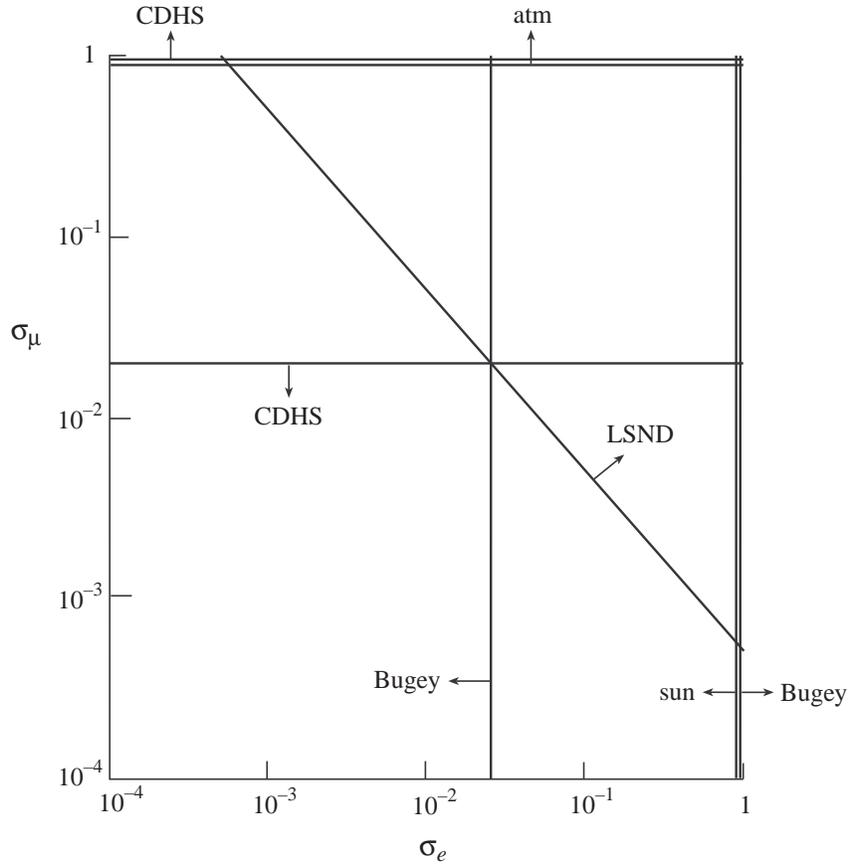}

\caption[]{Constraints on $\sigma_{\mu}$ and $\sigma_e$ defined in
Eq.~(\ref{sigma}) for the 2+2 model when $\delta m^2_{sbl}=1.7$~eV$^2$.
The LSND, Bugey, CDHS, solar and atmospheric constraints are obtained
by comparison of the appropriate data with Eqs.~(\ref{schwartz}),
(\ref{enotepairs}), (\ref{munotmupairs}), (\ref{sunboundpairs}), and
(\ref{atmboundpairs}), respectively.}
\end{figure}


\begin{figure}
\centering\leavevmode
\epsfxsize=4in\epsffile{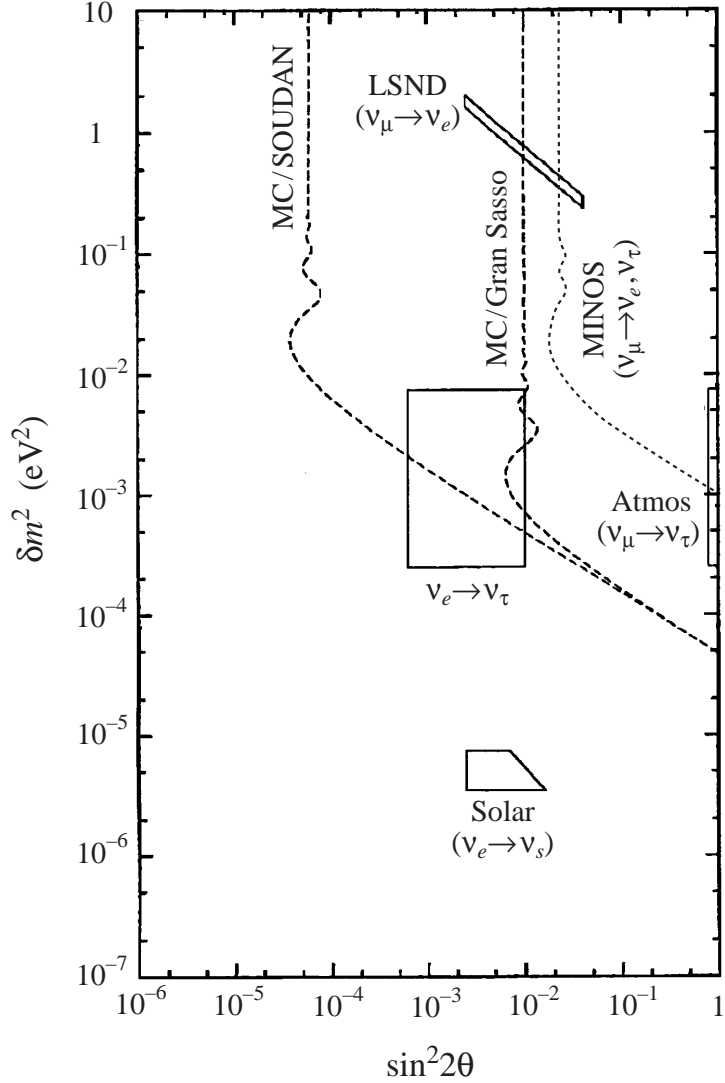}

\caption[]{Predicted region in the effective $\delta
m^2$-$\sin^22\theta$ parameter space for $\nu_e \rightarrow \nu_\tau$
oscillations in the four-neutrino model (solid rectangle), which is
determined by ${1\over4}$ of the
LSND $\nu_\mu \rightarrow \nu_e$ oscillation amplitude and the
atmospheric neutrino $\nu_\mu \rightarrow \nu_\tau$ oscillation
$\delta m^2$ scale. The dotted curves show the potential limits on
$\nu_\mu\rightarrow \nu_e, \nu_\tau$ oscillations from the MINOS
experiment~\cite{MINOS} and the dashed curves show the potential limits
on $\nu_e,\nu_\mu\rightarrow\nu_\tau$ oscillations that can be set by
neutrino beams from an intense muon source at
Fermilab~\cite{Geer} to detectors at the SOUDAN and GRAN SASSO sites for
muons with energy of 20~GeV. Also shown are the parameters for the solar
$\nu_e \rightarrow \nu_s$ small-angle MSW oscillation.}
\end{figure}

\end{document}